\documentclass[lettersize,journal]{IEEEtran}
\usepackage{comment}
\usepackage{calc,amsfonts,amssymb,amsmath,bm,url,color,theorem,graphicx,cite}
\usepackage{psfrag,subfigure,float}
\usepackage{multirow,subfigure,amsmath,epsfig,amsfonts,amssymb,psfig,graphics,psfrag,theorem,calc,subfigure,url,bm,cite}
\usepackage{stfloats,hyperref}  
\usepackage{algorithm,algorithmic}
\usepackage{diagbox} 
\usepackage{hhline}
\usepackage{caption}
\usepackage{bbding}
\usepackage{booktabs}
\usepackage{textcomp}
\usepackage{threeparttable}
\usepackage{graphicx}
\usepackage{multirow}
\usepackage{makecell}
\usepackage{booktabs}
\usepackage{fancyhdr}
\usepackage{tikz}
\usepackage[table]{xcolor}
\usetikzlibrary{quantikz}
\usepackage{mathtools}

\def\blue{\color{blue}}
\def\red{\color{red}}
\usepackage{colortbl}

\makeatletter
\def\multilimits@{\bgroup
	\Let@
	\restore@math@cr
	\default@tag
	\baselineskip\fontdimen10 \scriptfont\tw@
	\advance\baselineskip\fontdimen12 \scriptfont\tw@
	\lineskip\thr@@\fontdimen8 \scriptfont\thr@@
	\lineskiplimit\lineskip
	\vbox\bgroup\ialign\bgroup\hfil$\m@th\scriptstyle{##}$\hfil\crcr}
\def\Sb{_\multilimits@}
\def\endSb{\crcr\egroup\egroup\egroup}
\makeatother

{\begin{list}{}%
		{\setlength{\rightmargin}{0pc}%
			\setlength{\leftmargin}{1pc} }} %
	{\end{list}}

\newlength{\twidth}
\definecolor{orange}{RGB}{255,107,0}
\def\blue{\color{blue}}

\def\red{\color{red}}

\theorembodyfont{\rmfamily}

{\begin{list}{}{
			\settowidth{\labelwidth}{\mbox{\textnormal{#1}}}%
			\setlength{\leftmargin}{\labelwidth+\labelsep}}}%
	{\end{list}}

\newcommand\bA{\ensuremath{{\bm A}}}
\newcommand\bB{\ensuremath{{\bm B}}}
\newcommand\bC{\ensuremath{{\bm C}}}
\newcommand\bD{\ensuremath{{\bm D}}}
\newcommand\bE{\ensuremath{{\bm E}}}
\newcommand\bF{\ensuremath{{\bm F}}}
\newcommand\bG{\ensuremath{{\bm G}}}
\newcommand\bH{\ensuremath{{\bm H}}}

\newcommand\bM{\ensuremath{{\bm M}}}

\newcommand\bQ{\ensuremath{{\bm Q}}}

\newcommand\ba{\ensuremath{{\bm a}}}

\newcommand\bv{\ensuremath{{\bm v}}}

\newcommand\bx{\ensuremath{{\bm x}}}

\definecolor{orange}{RGB}{255,107,0}
\def\blue{\color{blue}}
\def\red{\color{red}}

    \fancypagestyle{IEEEtitlepagestyle}{
	\fancyhf{}
	\fancyhead[LE,RO]{\thepage}
	\fancyfoot[C]{\scriptsize © 2026 IEEE. Personal use of this material is permitted. Permission from IEEE must be obtained for all other uses, in any current or future media, including reprinting/republishing this material for advertising or promotional purposes, creating new collective works, for resale or redistribution to servers or lists, or reuse of any copyrighted component of this work in other works. This is the accepted version of the manuscript. The final, published version is available at IEEE Xplore via DOI: 10.1109/TGRS.2026.3691164.}
	
    }

\begin{document}

\title{Hyperspectral Anomaly Detection Using Einstein Fuzzy Computing and Quantum Neural Network}

\author{Chia-Hsiang Lin,~\IEEEmembership{Senior Member,~IEEE}, Si-Sheng Young,~\IEEEmembership{Student Member,~IEEE}, \\ and Reza Langari,~\IEEEmembership{Senior Member,~IEEE}

\thanks{This study was supported by the Emerging Young Scholar Program (namely, the 2030 Cross-Generation Young Scholars Program) of National Science and Technology Council (NSTC), Taiwan, under Grant NSTC 114-2628-E-006-002.
This research was also supported by the ``Quantum AI and Convex Algorithm for Optical Satellite Anomaly Detection'' under the University Academic Alliance in Taiwan (UAAT), funded by the Ministry of Education (MoE).
   We thank the National Center for Theoretical Sciences (NCTS) and the National Center for High-performance Computing (NCHC) for providing the computing resources.}

    \thanks{\textit{(Corresponding author: Chia-Hsiang Lin) (Chia-Hsiang Lin, and Si-Sheng Young contributed equally to this work.)}}
    \thanks{C.-H. Lin is with the Department of Electrical Engineering, National Cheng Kung University, Tainan, Taiwan (R.O.C.) (e-mail: chiahsiang.steven.lin@gmail.com).}
    \thanks{S.-S. Young is with the Institute of Computer and Communication Engineering, Department of Electrical Engineering, National Cheng Kung University, Tainan, Taiwan (R.O.C.) (e-mail: q38121509@gs.ncku.edu.tw).}
    \thanks{R. Langari is with the Department of Mechanical Engineering, Texas A\&M University at College Station, College Station, TX 77840 USA (e-mail: rlangri@tamu.edu).
    }

 }

\maketitle

\begin{abstract}
In the remote sensing (RS) field, hyperspectral imagery provides rich spectral information and facilitates numerous critical applications, such as material identification.
Among these applications, hyperspectral anomaly detection (HAD) aims to detect substances whose spectral characteristics deviate from background spectra, which are termed anomalies.
However, many widely used HAD algorithms in the RS community identify anomalies by relying on a ``background reconstruction'' strategy.
Furthermore, the lack of prior target hyperspectrum and real-world limitations collectively reduces the spectral discrepancy between anomaly and background, limiting the performance of mainstream detections.
By exploring the widely applicable fuzzy theory in the RS field, this study develops an unsupervised hybrid quantum-fuzzy multi-criteria decision framework (HyFuHAD) to detect anomalies from multiple perspectives.
In our HyFuHAD, each pixel is first fuzzified using multiple HAD-based membership functions (MFs), including morphological, geometrical, and statistical MFs, to obtain various types of fuzzy degrees.
Then, a multi-fuzzy-rule system, empowered by Einstein fuzzy computing, infers the classical fuzzy detection from these fuzzy degrees with sub-second-level computing.
The Einstein sum and product provide significantly smoother transitions compared to typical min-max-based fuzzy ``OR'' and ``AND'' during the fuzzy matching and inference steps, thereby enabling effective detections.  
Moreover, a lightweight quantum defuzzifier obtains the quantum fuzzy detection from fuzzy features derived from the proposed fuzzy feature aggregation network. 
Experiments demonstrate that our HyFuHAD algorithm achieves state-of-the-art performance by fusing the information from the quantum and classical detectors.
The demo code will be publicly available at \url{https://github.com/IHCLab/HyFuHAD}. 
The Supplementary Material can be found at \href{https://drive.google.com/file/d/1wcstj7Fbc5IeWjxOAaSsbAsQh03rKA0W/view?usp=drive_link}{Link}.
\end{abstract}


\section{Introduction}\label{sec: intro}
The rapid development of remote sensing (RS) technology has demonstrated remarkable potential in engineering and computer vision applications.
Nowadays, advanced hyperspectral imagery systems are capable of capturing broad-ranging electromagnetic (EM) reflectance across visible to near-infrared (NIR) or even middle-infrared (MIR) wavelengths, thereby forming high spectral resolution hyperspectral images (HSIs) \cite{8314827}. 
Even for the widely applicable multispectral imagery, recent RS literature \cite{COS2A} has demonstrated the feasibility of reconstructing AVIRIS-level HSIs from the Sentinel-2 counterparts using the Convex/Deep framework \cite{CODE,CODEIF,CODEMM}. 
Accordingly, HSIs become more and more popular in the RS fields and empower dozens of crucial RS tasks from low-level applications \cite{HyperKING,PRIME,DAEN} to high-level material identification and understanding\cite{QUEENG,CODEMM}, especially, hyperspectral anomaly detection (HAD) \cite{SuperRPCA,TGFA-AD, GTVLRR,li2019hyperspectral}.
HAD aims to identify spectral signatures lying in HSIs that significantly deviate from the background components, termed anomalies \cite{reviewer1,reviewer2,reviewer4}.
In contrast to hyperspectral target detection \cite{6678247}, spectral or spatial prior knowledge regarding anomalies is practically unavailable in the HAD task.
Under such constraints, mainstream HAD studies focus on ``background reconstruction'' and subsequently identify the residual as detection results, rather than directly evaluating spectral characteristics within target areas.
For example, conventional methods often assume that background components adhere to specific handcrafted properties, such as low-rank \cite{SuperRPCA,ADLR}, total variation \cite{LARTVAD}, and graph structure \cite{GTVLRR}, thereby formulating specialized optimization frameworks for HAD.   
Alternatively, deep learning (DL)-based methods rely on the so-called ``black-box'' models for background reconstruction \cite{PUNNet,GT-HAD,TAEF,TGFA-AD,reviewer5}.
Since background components typically dominate the overall hyperspectral structure, self-supervised deep networks are employed to facilitate more effective background reconstruction.
For example, an autonomous HAD network (Auto-AD) \cite{Auto-AD} reconstructs background components utilizing a fully convolutional neural network (CNN) integrated with an adaptive weighted loss function.
Building upon this, numerous DL-based HAD methods have been developed based on diverse network architectures \cite{RGAE,BockNet,GT-HAD,NL2Net,reviewer6}; the reconstruction strategy is also referred to as residual-based detectors.

Nevertheless, relying solely on residual-based detectors may encounter several challenges.
Specifically, due to the lack of prior information regarding anomalous targets \cite{reviewer3}, only handcrafted properties can be used for background extraction.
However, they often fail to model complex real-world scenarios, rendering optimization-based methods less effective.
By comparison, black-box networks may suffer from their superior reconstruction capability.
When the model overly recovers both background and anomaly components, the residual lacks informative structure to identify anomalies.   
On the other hand, HSIs are usually acquired with coarser spatial resolution compared to their multispectral or RGB counterparts due to hardware considerations \cite{CODEIF}.
This inherent restriction renders spectral mixtures of anomaly and background signatures (i.e., mixed-pixel phenomenon \cite{HiSUN}), indicating that anomaly may also be recovered during the reconstruction procedure, despite employing an effective reconstructor.
Furthermore, noise and outlier effects are also non-negligible matters \cite{RGAE} in practical scenarios, which hinder effective background reconstruction.
Consequently, the reconstruction residual alone provides limited information, preventing the mainstream HAD frameworks from reliable anomaly identification.

To enhance practical applicability, a straightforward yet effective solution is to incorporate additional information for further decision-making (DM) \cite{Rezabook}. 
Specifically, when anomalies are identified from multiple heterogeneous perspectives (which may include residual-based strategies), the resulting detection maps can be regarded as complementary components.
Subsequently, a specialized DM framework tailored for HAD can efficiently integrate these maps to yield an improved result.
However, typical DM approaches are heuristic-driven and may lack clear logical explanations \cite{li2015decision,xiang2022spectral}.
In contrast, within the fuzzy community, fuzzy theory-based multi-criteria DM (MCDM), known for its high flexibility and efficiency, has demonstrated remarkable success across numerous applications \cite{kahraman2015fuzzy}. 
Accordingly, this study, for the first time, develops an unsupervised fuzzy MCDM framework that exploits diverse anomaly characteristics to address the HAD task. 
To understand the design of our method, certain preliminaries of fuzzy theory needed are briefly reviewed in Section \ref{Preliminary}; and the proposed methods are detailed as follows.

In the proposed MCDM framework, we initially use computationally efficient fuzzy membership functions (MFs) rather than relying on existing HAD algorithms (mainly developed using residual strategy and may be less efficient) to obtain informative detection maps; this process is referred to as fuzzification in the fuzzy theory.
Each MF characterizes a linguistic condition (see Section \ref{Preliminary}), namely \textit{Under a Specific Property, Which Pixels Belong to the Anomaly?} 
In our framework, statistical, morphological, and geometrical properties are selected to construct the corresponding MF for HAD.
Among them, the geometrical MF associates with a geometrical unmixing process \cite{HyperCSI}, followed by a residual detection strategy.
In this sense, the proposed framework not only incorporates additional information but also preserves the strength of the residual detectors.
Following the fuzzification, we design a multi-logical fuzzy matching step based on fuzzy if-then rules to infer the result from the MFs, wherein the Einstein fuzzy operators implement the fuzzy logics \cite{Rezabook}, i.e., Einstein fuzzy computing.
These operators can provide more flexible and effective degree transitions compared to typical fuzzy ``AND'' and ``OR'' operators.
Finally, the defuzzification step, which comprises spatial-smoothness filtering and automatic energy-contrastive enhancement, is developed to yield the decision result, i.e., classical fuzzy detection (CFD).
Remarkably, the classical MCDM is unsupervised while achieving sub-second-level computing (see Section \ref{subsec: ablation}), outperforming most HAD methods in computational efficiency.

Furthermore, we proposed a novel fuzzy feature aggregation network to improve the generality of the proposed MCDM framework. 
In detail, the informative maps obtained from the diverse MFs, as well as their fuzzy complementary representation (see Section \ref{Preliminary}), are further fuzzified through trainable Gaussian MFs.
This deep fuzzification encodes the conventional fuzzy maps into the fuzzy feature space for complex real-world representation.
Subsequently, inspired by the concept of extended fuzzy sets \cite{extenedFSs}, trainable tokens are incorporated as degrees of hesitancy.
With these tokens, the fuzzy degrees of a pixel can represent not only the anomaly $a$ and background $b$ but also additional hesitancy $h$, hence improving flexibility and representation of fuzzy features.

In the defuzzification layer, we employ the quantum deep network (QUEEN) as a quantum-based defuzzifier due to its superiority in the RS area.
In fact, QUEENs have not only been advanced in the fuzzy systems community \cite{Quantum_DEFUZZIFIER} but also demonstrated their superiority in RS applications \cite{SQUARE_Mamba,QEDNet}.
The complex quantum space \cite{HyperQUEEN} provides a promising expressibility in feature aggregation and DM, while typically having a much smaller amount of parameters compared to conventional models.
For instance, a QUEEN-based deep image prior has been developed recently to address the challenging underdetermined blind source separation problem \cite{GQmu}.
These advantages collectively motivate us to conduct the quantum-based defuzzification for a decision result, i.e., quantum fuzzy detection (QFD); and the entire learning process is termed quantum MCDM.
Eventually, by fusing the information of the CFD and QFD, the proposed framework, named hybrid quantum-fuzzy MCDM (HyFuHAD), achieves state-of-the-art (SOTA) performance compared to numerous representative HAD benchmarks.
Moreover, the ablation study (see Section \ref{subsec: ablation}) demonstrates that the proposed quantum MCDM is necessary for enhancing the overall performance of HyFuHAD.
The main contributions of this study are outlined below:
\begin{enumerate}
    \item Although background reconstruction strategies enable mainstream HAD algorithms to achieve promising detection performance in certain scenarios, detecting anomalies solely from residual information suffers from limited generalization and reliability.  
    To enhance practical applicability, an effective approach is to incorporate additional information to produce heterogeneous detection maps and further develop a DM framework to obtain improved results.
    Accordingly, we develop a fuzzy-quantum-driven MCDM framework, named HyFuHAD, to achieve effective HAD performance.
    \item The proposed HyFuHAD first includes a sub-second-level computing classical MCDM. 
    Specifically, we obtain the heterogeneous maps by diverse computationally efficient MFs rather than existing HAD algorithms.
    Then, we tailor fuzzy logical if-then rules to infer the improved detection map, in which the Einstein fuzzy operators are employed to implement fuzzy logics for flexible degree transitions.   
    Notably, the classical MCDM is unsupervised and exhibits superior computational efficiency compared to the benchmark HAD algorithms.

    \item Furthermore, we propose a novel fuzzy feature aggregation network to learn deep fuzzy features from the heterogeneous maps.
    Specifically, trainable MFs first fuzzify these maps into deep fuzzy representations, while deep fuzzy if-then rules are subsequently employed to derive the corresponding fuzzy features for modeling complex real-world scenarios.
    Then, we employ the advanced quantum technique, QUEEN, as a quantum-based defuzzifier. 
    The quantum space offers high expressibility for feature aggregation and DM, while often remaining architecturally lighter than conventional models. 
    The resulting quantum-driven framework is referred to as quantum MCDM.

    \item By fusing information from quantum and classical MCDM, the proposed HyFuHAD achieves SOTA detection performance compared to 12 representative and latest HAD algorithms, across eight real-world scenarios.
    The ablation studies further substantiate the necessity of incorporating the Einstein fuzzy operators and advanced quantum MCDM.
    Furthermore, extensive case studies demonstrate that our HyFuHAD framework is generalizable and effective across diverse spatial scales and real-world applications (e.g., oil spill detection).
\end{enumerate}
The remainder of this study is summarized as follows.
In Section \ref{Preliminary}, we briefly recall the preliminaries of fuzzy sets, fuzzy logics, and fuzzy rule-based inference.
Then, Section \ref{Proposed} provides the details of the proposed HyFuHAD, including the designs of the overall framework, classical MCDM, and quantum MCDM.
Comprehensive experiments, including the datasets, experimental settings, qualitative and quantitative evaluations, as well as ablation, are summarized in Section \ref{sec: Experiment}.
Finally, our conclusion is presented in Section \ref{Conclusion}.

The frequently used notations throughout this study are defined as below.
$\mathbb{R}^H$, $\mathbb{R}^{H\times W}$, and $\mathbb{R}^{H\times W \times C}$ represent the $H$-dimensional Euclidean space, ($H \times W$)-dimensional real-valued matrix space, and ($H \times W \times C$)-dimensional real-valued 3-way tensor space, respectively.
Given a positive real-valued integer $L$, we define $\mathcal{I}_L\triangleq\{1,2,\cdots,L\}$.
Given a classical set $\Omega$, its cardinality is denoted as $|\Omega|$.
Given a vector $\bv$, we use $\bv_i$ to denote the $i$th entry of $\bv$.
Given a matrix $\bM$, the transpose, inverse, and determinant of $\bM$ are denoted as $\bM^T$, $\bM^{-1}$, and $|\bM|$, respectively.
$[\mathbf{Z}]_{k_1,k_2,\cdots,k_N}$ denotes the (${k_1,k_2,\cdots,k_N}$)th entry of the $N$-way tensor $\mathbf{Z}$.
Given two $N$-way tensors $\mathbf{X}$ and $\mathbf{Y}$, their Hadamard product is denoted as $\mathbf{Z}=\mathbf{X}\odot\mathbf{Y}$, with $[\mathbf{Z}]_{k_1,k_2,\cdots,k_N}=[\mathbf{X}]_{k_1,k_2,\cdots,k_N} \times [\mathbf{Y}]_{k_1,k_2,\cdots,k_N}$.
Given two real-valued scalar $x,y\in[0,1]$, the Einstein product $x\wedge y=\frac{xy}{2-(x+y-xy)}$ and the Einstein sum $x\vee y=\frac{x+y}{1+xy}$.
Finally, $\| \cdot\|_2$ and $\| \cdot\|_F$ are the $\ell_2$- norm and Frobenius norm, respectively.
%


\section{Preliminary: From Fuzzy Set, Fuzzy Logic, to Fuzzy Rule-based Inference}\label{Preliminary}
In this section, we recall the essential concepts of fuzzy MCDM for a better understanding of the proposed HAD technique.
First, a classical set $C$ is considered as a collection of elements with sharp degrees; for example, given an element or a sample $x_k$, the degree is either 1 (i.e., $x_k\in C$) or 0 (i.e., $x_k\notin C$). 
In contrast, a fuzzy set $A$ is typically characterized by smooth degrees and can be expressed as
\begin{align}\label{fuzzyset}
    A = \left\{ \left( x_k, \mu_A(x_k) \right) \,\middle|\, x_k \in X \right\},
\end{align}
where $\mu_A: X\rightarrow[0,1]$ denotes the MF describing whether $x_k$ is entirely not, partially, or completely in $A$. 
Also, $X$ is the domain of concern and is usually called ``\textit{universe of discourse}'' in fuzzy systems \cite{Rezabook}.
Due to the flexibility of fuzzy sets, they excel at handling imprecise information for various applications.
To illustrate, one assigns a set regarding ``YOUNG'' to look for people who appear to be between 20 to 30 years old, and assumes the universe of discourse $X:=[0,+\infty]$.
Accordingly, the degree of classical set $C:=[20,30]$ fails to represent the marginal scenarios, for example, 19 or 31 years old (${19,31} \notin C$), even though these samples appear indistinguishable from those within the target range.
Whereas, the fuzzy set $A$ allows the samples with imprecise information or boundary, like 19 or 31 years old, to partially belong to it by building a suitable MF, for example,
\begin{equation*}
\mu_A(x)
=
\begin{cases}
\exp\left({- \frac{(x - 20)^2}{2 (1.5)^2}}\right) &\text{when} ~~  0\leq x <20,
\\
~~~~~1 &\text{when} ~~ 20\leq x \leq 30,
\\
\exp\left({- \frac{(x - 30)^2}{2 (1.5)^2}}\right) &\text{when} ~~   x >30,
\end{cases}
\end{equation*}
thereby addressing challenging real-world applications.

Based on the aforementioned concepts, one can realize that the core advantages of fuzzy sets come from their ability to model a gradual transition for ambiguous scenarios.
Thus, how to determine an appropriate MF becomes a vital matter.
Typically, MFs are built by several approaches, such as tuning through a trial-and-error process empirically, constructing from big data, or learning using the feedback/objectives from a fuzzy system (e.g., FNNs), etc.
The widely used approach is to parameterize the functions using a small number of parameters, for example, Gaussian, Sigmoidal, and trapezoid MFs \cite{Rezabook}. 
On the other hand, we remark that quantum systems can also be interpreted using quantum fuzzy sets (QFS) representation \cite{Quantum_DEFUZZIFIER}.
To understand this, we refer the interested readers to \cite[Section II-A]{HyperQUEEN} for the quantum preliminaries, and we introduce the concept of QFS hereinafter. 
A quantum state at a specific moment in an arbitrary $n$-quantum bit (qubit) circuit can be expressed as $\ket{\psi}=\sum_{i=0}^{2^n-1} p_i \ket{\psi}_i$, where $\ket{\psi}_0,\dots,\ket{\psi}_{2^n-1}$ are the computational basis states; and $p_0,\dots,p_{2^n-1} \in \mathbb{C}$ satisfying $\sum_{i=0}^{2^n-1} |p_i|^2=1$ are the probability amplitudes.
Accordingly, one can easily interpreted an $n$-qubit state  $\ket{\psi}$ using its fuzzy set representation $\psi$, i.e., 
\begin{align}\label{QFS}
    \psi = \left\{ \left( \ket{\psi}_i, \mu_{\psi}(\ket{\psi}_i) \right) \,\middle|\, i \in \{0,\cdots,{2^n-1}\} \right\},
\end{align}
where $\mu_{\psi}(\ket{\psi}_i)=|p_i|^2$.
Such a closed relation motivates researchers to advance the parameterized quantum circuits (i.e., QUEENs) to handle the uncertainties and fuzziness for complex-valued fuzzy systems, such as sentiment and sarcasm detection \cite{Quantum_DEFUZZIFIER,TFS_CFN}.
More theoretical analysis on quantum and fuzzy logics can be found in \cite{molecules26195987,melnichenko2007quantum}.
In summary, parameterized and differentiable functions are typically adapted as traditional MFs, and QUEENs are employed as the complex-valued fuzzifiers in advanced fuzzy systems.

After we have acknowledged the concept of fuzzy sets and how to build their MFs, we proceed to introduce fuzzy logic as follows. 
Similar to the fundamental operations for classical sets, union $\cup$, intersection $\cap$, and complement $C^{c}$, these operations are also generalized in fuzzy systems to logically denote ``AND'', ``OR'', and ``NOT'', respectively, for fuzzy rules.
In detail, given two arbitrary fuzzy sets $A$ (with MF $\mu_{A}$) and $B$ (with MF $\mu_{B}$), and a sample from the universe of discourse $x_k\in X$, the fuzzy union, fuzzy intersection, and fuzzy complement are often defined using the maximum operator, minimum operator, and the difference between one and its MF value, respectively, i.e.,
\begin{align}
    \mu_{A\cup B} (x):=&\max\{\mu_{A}(x),\mu_{B}(x)\}, \notag
    \\
    \mu_{A\cap B} (x):=&\min\{\mu_{A}(x),\mu_{B}(x)\},\notag
    \\
    \mu_{A^{c}} (x):=&1-\mu_{A}(x).
\end{align}
Remarkably, in fuzzy systems, the definitions of these fuzzy operations are not unique.
The collection of candidate fuzzy conjunctions (i.e., intersection) is termed ``triangular norms''; and the collection of candidate fuzzy disjunctions (i.e., union) is named ``triangular conorms''.
Typically, triangular norms $t: [0,1]\times[0,1]\rightarrow[0,1]$ and triangular conorms $s: [0,1]\times[0,1]\rightarrow[0,1]$ form a dual pair satisfying the condition, i.e.,
\begin{align}
    1-t(\mu_{A}(x),\mu_{B}(x))=s(1-\mu_{A}(x),1-\mu_{B}(x)),
\end{align}
which is similar to De Morgan's laws in classical logic theory, i.e., $(A\cap B)^{c}=A^{c}\cup B^{c}$, when we consider the operator ``$1-y$'' as the fuzzy complement. %
Although there are so many choices for the dual pairs of triangular norms and conorms, both of them must satisfy the fundamental properties of the set of dual axioms, such as monotonicity, commutativity, and associativity; and we refer interested readers to \cite[Chapter 3]{Rezabook} for a comprehensive introduction.
The widely used dual pairs include minimum and maximum operators, algebraic product and algebraic sum, as well as the Einstein product and Einstein sum, and so on.
Among these pairs, the minimum and maximum operators provide substantially crisp and less informative matching results when two degrees are close to each other.
When two degrees are significantly different, these operators fail to provide gradual transitions.
In contrast, both the algebraic and Einstein operators exhibit significantly more gradual characteristics compared to the minimum and maximum operators, in which the Einstein operators further provide smoother transitions for extreme degrees (i.e., values close to 0 or 1).
Having these foundations for basic fuzzy logic operators, we can now elaborate on the fuzzy-based DM.

The structure of a basic fuzzy if-then rule contains two components, i.e., if-part and then-part, and is usually denoted as IF \textless \textbf{antecedent} \textgreater\ THEN \textless \textbf{consequent} \textgreater\ in fuzzy systems.
For example, IF \textless the temperature is \textit{High} \textgreater\ THEN \textless the weather is \textit{Hot}\textgreater, in which the terms to describe the fuzzy concepts (i.e., \textit{High} and \textit{Hot}) are called ``linguistic variables''.
To represent the fuzziness of linguistic variables, an antecedent is typically characterized by an MF of a specific fuzzy set; for example, given a temperature $T$ and a MF $\mu_{\textit{High}}(\cdot)$ describing the linguistic variable \textit{High}, one can obtain the degree of antecedent $\mu_{\textit{High}}(T)$.
Beyond the single antecedent scenario, a fuzzy rule can also comprise $n$ antecedents that are represented by $n$ fuzzy sets (denoted as $A_1,\cdots, A_n$) to better model the more complex scenarios.
For instance, IF \textless $x_1$ belongs to $A_1$ \textgreater\ AND \textless $x_2$ belongs to $A_2$\textgreater\ AND $\cdots$ AND \textless $x_n$ belongs to $A_n$\textgreater\ THEN \textless the weather is \textit{Comfortable} \textgreater\ .
To this end, the fuzzy conjunction operation (e.g., the minimum operator) is used to match the degree of multiple antecedents, i.e., 
\[
\textit{Matching Degree}=\min\{ \mu_{A_1}(x_1), \mu_{A_2}(x_2),\cdots, \mu_{A_n}(x_n)\}.
\]
The above process is known as the ``fuzzy matching'' (FM).
Following the FM, ``fuzzy inference'' (FI), typically employing the clipping and scaling methods, is conducted to derive a fuzzy conclusion for the consequent.
Specifically, a typical consequent can be classified into three major types: 1) crisp consequent, assigning a symbolic or numeric value, 2) fuzzy consequent, assigning a fuzzy set, or 3) functional consequent, executing a predefined functional operation.
Taking the second consequent as an example, we can assign a fuzzy set characterized by its MF $\mu_{\textit{C}}(\cdot)$ to the linguistic variable \textit{Comfortable}, then the FI step can be expressed as $\min\{\mu_{\textit{C}}(y),\textit{Matching Degree}\}$ for clipping method; and $\mu_{\textit{C}}(y)\times \textit{Matching Degree}$ for scaling method.
Also, $y$ is a specific sample belonging to the universe of
discourse of $\mu_{\textit{C}}$.
Accordingly, after the FI step, a lower matching degree leads to a more significant suppression of the consequent MFs.

So far, we have acknowledged how to infer a fuzzy conclusion from a single fuzzy rule (even with multiple antecedents).
Subsequently, for those fuzzy systems containing multiple and potentially overlapping fuzzy rules, ``fuzzy combination'' (FC) is required to integrate the multiple inferred fuzzy conclusions.
A commonly used approach is to perform the fuzzy disjunction operation (e.g., maximum operator) to aggregate multiple inferred fuzzy conclusions.
Eventually, the process converting the aggregated fuzzy conclusions from the FC step into a crisp decision is known as ``defuzzification''.
In a typical fuzzy rule-based inference, defuzzification can often be accomplished by several strategies, e.g., the Mean of Maximum (MOM) or Center of Area (COA) methods  \cite{Rezabook}.
It is worth noting that the defuzzification step is optional and only needed in applications that necessitate a non-fuzzy decision.
With the above foundational concepts for fuzzy systems, we can proceed to the proposed HyFuHAD.
In the sequel, boldface notation is adopted to represent the proposed fuzzy rules; for instance, \textbf{R1} indicates the first fuzzy rule.

\section{Proposed Method}\label{Proposed}
The structure of this section is organized as follows for a clear understanding.
First, we will begin with the design of the overall framework in Section \ref{subsec: overall} to construct the big picture.
Later, we elaborate on the design of the classical MCDM in Section \ref{subsec: CFD}.
After that, we introduce the design of our quantum MCDM, including the architecture of the fuzzy feature aggregation network, quantum defuzzifier, and the loss function, in Section \ref{subsec: QFD}.
\subsection{Overall framework}\label{subsec: overall}
\begin{figure}[t]
    \centering
    \includegraphics[width=1\linewidth]{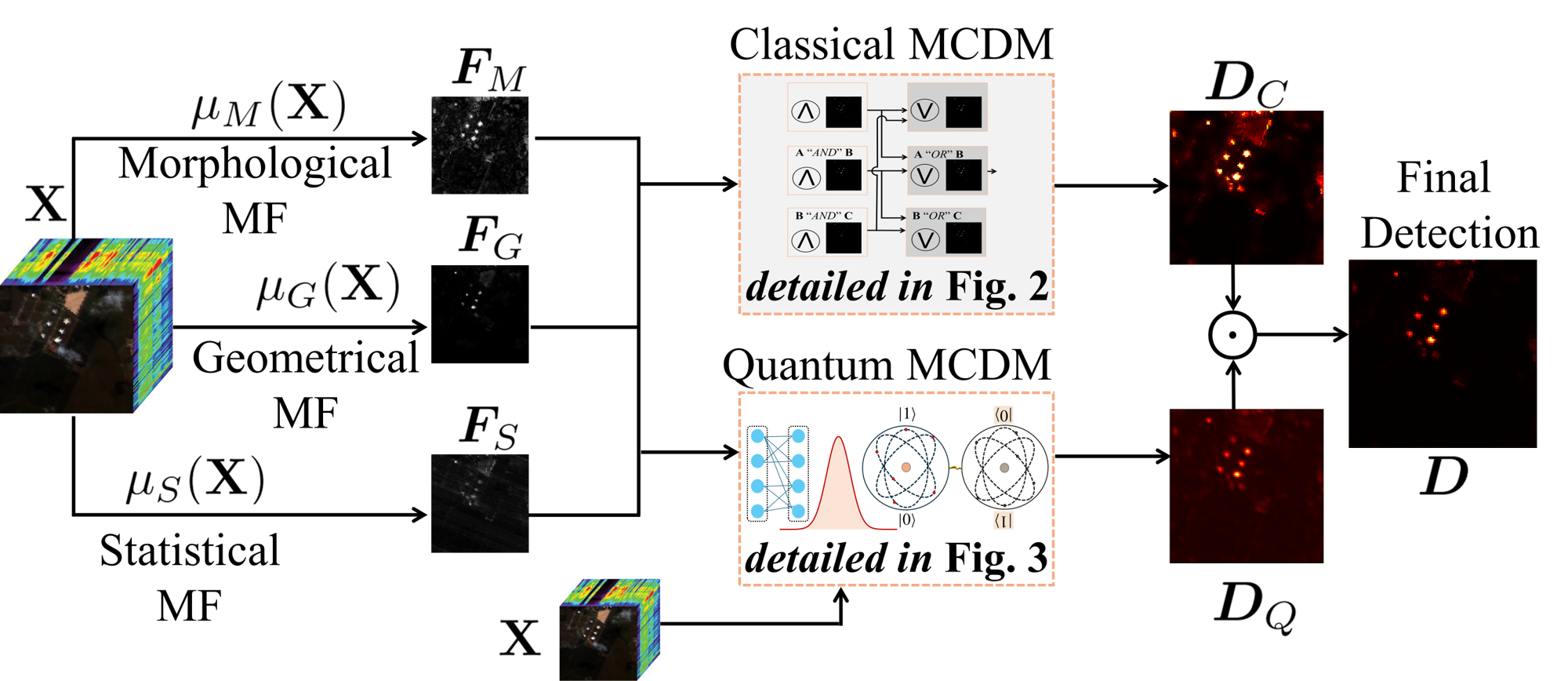}
    \vspace{-0.6cm}
    \caption{Overview of the proposed HyFuHAD.
    First, input HSI $\mathbf{X}$ is transformed into fuzzy degree maps (i.e., $\bF_M$, $\bF_G$, and $\bF_S$) using multiple HAD-based MFs, namely fuzzification.
    After the fuzzification, we perform both classical and quantum MCDMs to infer the CFD $\bD_C$ and QFD $\bD_Q$, respectively. 
    Eventually, we obtain the final detection $\bD$ by integrating the CFD and QFD.
    More details are presented in Section \ref{subsec: overall}.}\label{fig: overall_famework}
    \vspace{-0.3cm}
\end{figure}
%
Given a HSI with $C$ channels and $L=HW$ pixels,  denoted as $\mathbf{X}\in\mathbb{R}^{H\times W \times C}$, our proposed framework (see Figure \ref{fig: overall_famework}) begins with constructing multiple fuzzy sets that are linguistically described as, ``\textit{Under Specific Properties, Which Pixels Belong to the Anomaly}''.
That is, given a sample (i.e., pixel) from the universe of discourse (i.e., input HSI), its fuzzy degree, characterized by the MF, explicitly represents its possibility of being an anomaly.
In light of the diversity of anomalies in real-world scenarios, modeling the anomalies using multiple fuzzy sets that represent diverse properties is necessary.
Furthermore, to fulfill practical applications, we require a highly flexible, computationally efficient, as well as unsupervised fuzzification step.
Accordingly, conventional MFs, such as Gaussian, Sigmoidal, and trapezoid MFs, are unsuitable for the proposed framework.
To realize these advantages, we customize the fuzzification step using three diverse HAD-based MFs, namely, morphological, geometrical, and statistical, and their details are presented as follows.
%
%
%
%

The first one is morphological MF $\mu_M(\cdot)$ implemented by morphological attribute opening and closing operators \cite{Morphological_2}.
In the morphological attribute filter (MAF), the openings operator $\mathcal{O}(\cdot)$ and closings operator $\mathcal{C}(\cdot)$ are used to remove the brighter and darker \textit{connected components} (CC), respectively, that are smaller than a predefined structure element \cite{Morphological_1}.
Here, CCs refer to groups of pixels that satisfy the connectivity rules (e.g., four-connected and eight-connected) in a binary image \cite{Connect_components}, which can be intuitively considered as adjacent regions.
For example, given a grayscale image $\bG\in [0,1]^{H\times W}$ and a threshold $\tau\in [0,1]$, one can binarize the grayscale image by 
\[
[\widehat\bG]_{i,j} = 
\begin{cases}
1, & \text{if } [\bG]_{i,j} \geq \tau, \\
0, & \text{if } [\bG]_{i,j} < \tau.
\end{cases}
\]
Then, the opening operator is used to suppress the grayscale value of those CCs with a boolean value of 1.
We refer interested readers to \cite{Morphological_1, Morphological_2} for more information on MAFs.
Since the anomalies are typically connected small targets (i.e., the morphological property), we use MAFs to efficiently capture their spatial structure by the following steps.
First, for computational efficiency, we project the HSI into the $P$-dimensional affine subspace ($P\ll C$) using signal subspace identification, followed by a standardization to obtain the principal components $\mathbf{P}\in\mathbb{R}^{H\times W \times P}$.
With careful observation, we find that the 3-dimensional affine subspace already preserves 99\% of the principal components for the target HSI.
Further enlarging $P$ may yield only marginal improvements (or introduce additional noise) while increasing the computational burden.
To realize real-time computing, we consider $P:=3$ a default setting for the morphological MF.
Subsequently, we apply the opening and closing operators on each principal component $[\mathbf{P}]_{:,:,i}~,~\forall i\in\mathcal{I}_P$ to capture the spatial structure of anomalies.  
Consequently, we obtain the morphological degree map $\mu_M(\mathbf{X}) =\bF_M\in\mathbb{R}^{H\times W}$ by
\begin{align}\label{morphological_MF}
   \bF_M=\sum_{i=1}^{P} |[\mathbf{P}]_{:,:,i} - \mathcal{O}([\mathbf{P}]_{:,:,i})| + |[\mathbf{P}]_{:,:,i} - \mathcal{C}([\mathbf{P}]_{:,:,i})|.
\end{align}

Next, we proceed to the design of the geometrical MF hereinafter.
To illustrate, we need to remark that blind source separation (BSS) is often used to estimate the $N$ pure spectral signatures $\bE \in \mathbb{R}^{C\times N}$ (i.e., columns of $\bE$) and their corresponding spatial proportion (i.e., abundance) $\mathbf{A} \in \mathbb{R}^{H\times W \times N}$ from a given HSI in RS technology \cite{HiSUN}.
In particular, convex geometry-based BSS methods, such as hyperplane-based Craig simplex identification (HyperCSI) \cite{HyperCSI}, have been theoretically proven to recover the true signatures and abundances under mild conditions \cite{7107995}.
%
%
Thus, assuming that anomalies are those substances with distinct spectral attributes (have a high possibility to have pure spectral signatures), then their abundance can be efficiently recovered using HyperCSI, namely, the geometrical property.
Therefore, we obtain the geometrical degrees map $\mu_G(\mathbf{X})=\bF_G\in\mathbb{R}^{H\times W}$ via the following steps.
First, we perform HyperCSI on the input HSI (with $N:=4$) \cite{TGFA-AD} to obtain $\mathbf{A}$, then the abundances are fused into the geometrical degrees map by
\begin{align}\label{geometrical_MF}
   \bF_G=\left(\sum_{i=1}^{N} a_i[\mathbf{A}]_{:,:,i}\right)\odot \bM,
\end{align}
where $a_i$ is the $i$th entry of the scene-adaptive linear projection vector (LPV) $\ba\in\mathbb{R}^{N}$, and $\bM=\frac{1}{C}\sum_{k=1}^C |\mathbf{X}-\widetilde{\mathbf{X}}|\in\mathbb{R}^{H\times W}$ is the guidance map inspired by conventional residual-based detectors \cite{TAEF,Auto-AD}.
Since the LPV is learned (will be detailed below), it provides adaptivity across scenarios and addresses the challenging model-order selection \cite{MDL} issue.
Specifically, although HyperCSI can only recover the abundance maps with $N:=4$ preliminarily, the LPV assigns the highest weight to the abundance that most closely aligns with the reference maps.
This advantage enables an effective fusion with a fixed model order.
Accordingly, the selection of model-order $N$ plays a relatively minor role for implementations.
Despite this, we observe that an insufficient model order combines the spatial distributions of various materials into a single abundance map, while an overly large one scatters the informative distributions across multiple abundances, both of which hinder LPV learning.
We empirically found that $ N=4$ is the minimum order sufficient to yield promising degree maps and generally applicable across diverse scenarios.
On the other hand, $\widetilde{\mathbf{X}}$ is the filtered HSI using the Gaussian filter with $\sigma:=5$.
Specifically, we use the guidance map to guide the LPV learning process due to an observation, i.e., an appropriate linear combination of abundance indeed provides graduate transitions for HAD-based fuzzification.
Therefore, we explicitly formulate this property as the following optimization problem \cite{TGFA-AD},
\begin{align}
\label{equ:closed form of learning}
\ba
&:=
\arg~\min_{{\widetilde{\ba}}\in\mathbb{R}^N }~\frac{1}{2}\|\bM-\sum_{i=1}^{N} \widetilde{a}_i[\mathbf{A}]_{:,:,i}\|_{F}^2.
\end{align}
Since the problem is convex, the optimal solution can be obtained using a gradient based approach.
At this point, we have completed the design of the geometrical MF.

Consequently, we present the implementation of statistical MF based on the Mahalanobis distance \cite{LRX}.
In the HAD field, the background pixels are typically assumed to follow a $C$-dimensional multivariate normal distribution \cite{multivariate}, whose probability density function, i.e.,
\[
p(\bx) = \frac{1}{(2\pi)^{C/2} |\bm{\Sigma}|^{1/2}} 
\exp\left( -\frac{1}{2} (\bx - \bm{\mu})^T \bm{\Sigma}^{-1} (\bx - \bm{\mu}) \right),
\]
where $\bm{\Sigma}\in\mathbb{R}^{C \times C}$ and $\bm{\mu}\in\mathbb{R}^C$ denote the covariance matrix and mean vector of HSI, respectively. 
However, the anomaly pixels usually deviate from the distribution, resulting in a significant distinction compared to background pixels, namely, the statistical property.
Hence, the statistical fuzzy degree map $\bF_S=\mu_S(\mathbf{X})$ can be estimated using the minimum log-likelihood estimation, 
\begin{align}
    [\bF_S]_{i,j}=([\mathbf{X}]_{i,j,:} - \bm{\mu})^T \bm{\Sigma}^{-1} ([\mathbf{X}]_{i,j,:} - \bm{\mu}),
\end{align}
which is exactly the Mahalanobis distance.
With the completion of the fuzzification steps, we can infer the CFD $\bD_C$ and QFD $\bD_Q$ by
\begin{align}
    \bD_C&=\text{C-MCDM}(\bF_M,\bF_G,\bF_S),\notag
    \\
    \bD_Q&=\text{Q-MCDM}(\bF_M,\bF_G,\bF_S,
    \mathbf{X}),
\end{align}
where $\text{C-MCDM}(\cdot)$ and $\text{Q-MCDM}(\cdot)$ denote the entire function of the proposed classical and quantum MCDM, respectively.
By fusing CFD and QFD, the final detection map, $\bD=\bD_C\odot\bD_Q$, accomplishes superior detections across various scenarios (see Section \ref{sec: Experiment}).
This multiplicative fusion scheme facilitates superior qualitative and quantitative performance compared with more sophisticated fusion strategies.
Specifically, we compare the self-supervised multi-scale CNN and Wavelet-based fusion strategies with the selected Hadamard product-based integration.
The self-supervised multi-scale CNN demonstrates scenario-dependent performance, potentially because of the absence of supervised guidance during optimization. 
The Wavelet fusion scheme reveals relatively limited background suppression capability.
In contrast, the multiplicative fusion scheme consistently achieves effective performance across diverse scenarios (see Section \ref{sec: Experiment}).
Accordingly, the simple yet effective Hadamard product is employed to derive the final detection map.
At this point, we have completed the introduction of the proposed HyFuHAD, and the remaining details for classical MCDM and quantum MCDM will be presented in the following Sections.  
\subsection{Classical Multi-Criteria Decision Making}\label{subsec: CFD}
\begin{figure}[t]
    \centering
    \includegraphics[width=1\linewidth]{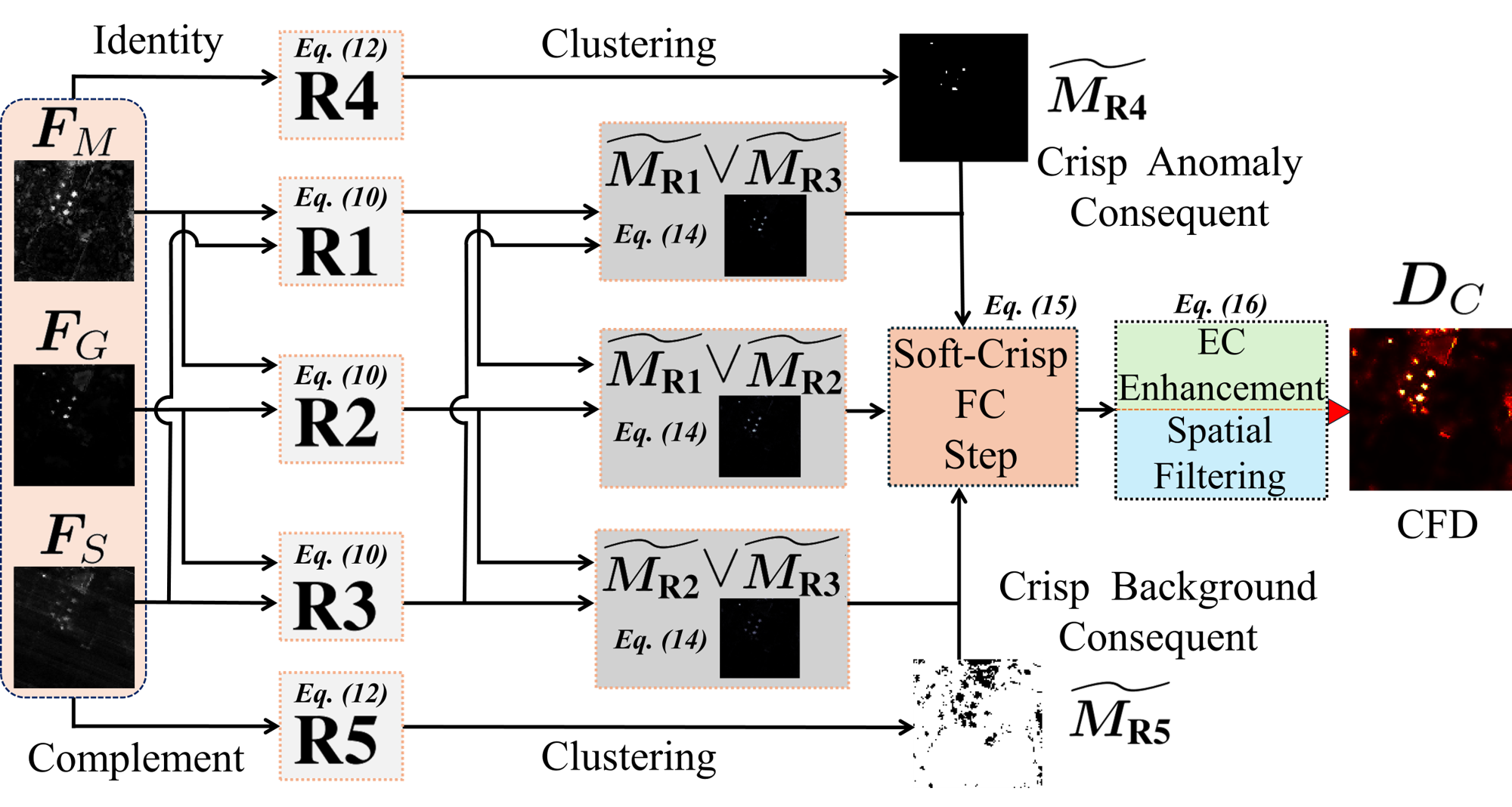}
     \vspace{-0.5cm}
    \caption{Overall flowchart of the proposed classical MCDM.
    Further details of the corresponding implementations are presented in Section \ref{subsec: CFD}.
    }\label{fig:classical_Fuzzy_detection}
    \vspace{-0.3cm}
\end{figure}
The proposed classical MCDM, as graphically illustrated in Figure \ref{fig:classical_Fuzzy_detection}, comprises fuzzy matching (FM), fuzzy inference (FI), fuzzy combination (FC), and defuzzification steps.
We refer readers to Section \ref{Preliminary} for the basic introduction of each step for a better understanding.
Initially, in the FM step, we develop five fuzzy rules (denoted as \textbf{R1}-\textbf{R5}) and match the fuzzy rules using fuzzy logic.
Specifically, we first assume that a pixel satisfying two HAD properties (e.g., geometrical and statistical properties) should be considered as an anomaly, resulting in the following fuzzy rules \textbf{R1}-\textbf{R3}, i.e.,
\begin{align}\label{equ: five_fuzzy_rule}
\textbf{R1:}&~\text{IF} ~[\bF_M]_{i,j}~ \text{AND}~ [\bF_S]_{i,j} ~\text{is High} ~\text{THEN}~ [\mathbf{X}]_{i,j,:}\in \mathcal{A},\notag
\\
\textbf{R2:}&~\text{IF} ~[\bF_M]_{i,j}~ \text{AND}~ [\bF_G]_{i,j} ~\text{is High} ~\text{THEN}~ [\mathbf{X}]_{i,j,:}\in \mathcal{A},\notag
\\
\textbf{R3:}&~\text{IF} ~[\bF_G]_{i,j}~ \text{AND}~ [\bF_S]_{i,j} ~\text{is High} ~\text{THEN}~ [\mathbf{X}]_{i,j,:}\in \mathcal{A}.
\end{align}
where $\mathcal{A}$ denotes the collection of all anomaly pixels within $\mathbf{X}$.
%
Due to the absence of reference anomaly maps, we intuitively consider the MF of $\mathcal{A}$ (denoted as $\mu_{\mathcal{A}}$) as an identity mapping, i.e., crisp consequent (see Section \ref{Preliminary}). 
To match the degrees for ``AND'' antecedents within the rules in \eqref{equ: five_fuzzy_rule}, we adapt the Einstein product as the fuzzy conjunction operator.
As introduced in the third paragraph of Section \ref{Preliminary}, the widely utilized minimum operator fails to provide gradual matching behavior.
When two degrees from various MFs differ significantly, the minimum operator results in a relatively crisp matching and fails to produce additional information, as the resulting degree is originally obtained from a single MF. 
In contrast, the Einstein product \cite{Fuzzy_Einstein} facilitates significantly smoother transitions for degree matching over the entire valid domain of fuzzy degrees.
By employing the Einstein computing, we can effectively prevent over-suppression of low-degree regions while providing smooth transitions.
Thus, the FM process can be expressed as
%
\begin{align}\label{equ: matching_R1_R3}
[\bM_{\textbf{R1}}]_{i,j}&=[\bF_M]_{i,j}\wedge[\bF_S]_{i,j},\notag
\\
[\bM_{\textbf{R2}}]_{i,j}&=[\bF_M]_{i,j}\wedge[\bF_G]_{i,j},\notag
\\
[\bM_{\textbf{R3}}]_{i,j}&=[\bF_G]_{i,j}\wedge[\bF_S]_{i,j},
\end{align}
where $\bM_{\textbf{R}k}\in\mathbb{R}^{H\times W}$ denotes the matching degree map of the $k$th rule.
As for the FI step for \textbf{R1}-\textbf{R3}, we adapt the scaling method (see Section \ref{Preliminary}) to infer the fuzzy conclusion for the $k$th rule $[\widetilde{\bM_{\textbf{R}k}}]$, namely, $[\widetilde{\bM_{\textbf{R}k}}]_{i,j}=\mu_{\mathcal{A}}([\mathbf{X}]_{i,j})\times[\bM_{\textbf{R}k}]_{i,j},~\forall k \in \mathcal{I}_3$.

Subsequently, we elaborate on the FI step for \textbf{R4}-\textbf{R5} hereinafter.
In addition to \textbf{R1}-\textbf{R3}, we further observe that if a pixel satisfies or violates all HAD properties simultaneously, then we hold a stronger confidence to identify it as an anomaly or background, respectively.
Therefore, these observations inspire us to construct the following fuzzy rule, 
\begin{align}\label{equ: five_fuzzy_rule2}
\textbf{R4:}&~\text{IF} ~[\bF_M]_{i,j}~ \text{AND}~ [\bF_G]_{i,j} ~ \text{AND}~ [\bF_S]_{i,j} ~\text{is High} \notag
\\
&\text{THEN}~ [\mathbf{X}]_{i,j,:}\in \mathcal{A},\notag
\\
\textbf{R5:}&~\text{IF} ~[\bF_M^c]_{i,j}~ \text{AND}~ [\bF_G^c]_{i,j} ~ \text{AND}~ [\bF_S^c]_{i,j} ~\text{is High} \notag
\\
&\text{THEN}~ [\mathbf{X}]_{i,j,:}\in \mathcal{B},
\end{align}
where $\mathcal{B}$ is the collection of all background pixels within $\mathbf{X}$, whose MF $\mu_{\mathcal{B}}$ is defined as the same as $\mu_{\mathcal{A}}$ for a similar reason.
According to \textbf{R4}-\textbf{R5}, their matching degree maps can be naturally obtained by
\begin{align}\label{equ: matching_R1_R3}
[\bM_{\textbf{R4}}]_{i,j}&\!=\![\bF_M]_{i,j}\!\wedge\![\bF_G]_{i,j}\!\wedge\![\bF_S]_{i,j},\notag
\\
[\bM_{\textbf{R5}}]_{i,j}&\!=\!(1\!-\![\bF_M]_{i,j})\!\wedge\!(1\!-\![\bF_G]_{i,j})\!\wedge\!(1\!-\![\bF_S]_{i,j}).
\end{align}
After the fuzzy matching, we adapt the functional consequent for the FI step of \textbf{R4}-\textbf{R5}.
In detail, we first perform the scaling method to obtain the preliminary fuzzy conclusions, then binarize them using dynamic thresholding (e.g., K-means \cite{kmeans}) due to their high certainty grounded in the three HAD properties, i.e., $\widetilde{\bM_{\textbf{R}k}}=\text{K-means}(\bM_{\textbf{R}k}),~\forall k\in\{4,5\}$.  
From the above steps, we have obtained the fuzzy conclusions, $\widetilde{\bM_{\textbf{R}k}},~\forall k\in\mathcal{I}_5$, for all fuzzy rules in \eqref{equ: five_fuzzy_rule} and \eqref{equ: five_fuzzy_rule2}.

Now, we will proceed to the FC step to integrate the multiple fuzzy conclusions.
First, for the three soft fuzzy conclusions (i.e., $\widetilde{\bM_{\textbf{R1}}}$, $\widetilde{\bM_{\textbf{R2}}}$, and $\widetilde{\bM_{\textbf{R3}}}$), we adapt the cross-integration (see Figure \ref{fig:classical_Fuzzy_detection}) implemented by fuzzy disjunction, i.e., 
\begin{align}\label{equ: cross_disjunction}
[\bC_{1}]_{i,j}&=[\widetilde{\bM_{\textbf{R1}}}]_{i,j}\vee[\widetilde{\bM_{\textbf{R3}}}]_{i,j},\notag
\\
[\bC_{2}]_{i,j}&=[\widetilde{\bM_{\textbf{R1}}}]_{i,j}\vee[\widetilde{\bM_{\textbf{R2}}}]_{i,j},\notag
\\
[\bC_{3}]_{i,j}&=[\widetilde{\bM_{\textbf{R2}}}]_{i,j}\vee[\widetilde{\bM_{\textbf{R3}}}]_{i,j},
\end{align}
where the fuzzy disjunction is realized via Einstein sum to mitigate the over-enhancement in high-degree regions, thereby ensuring stable inference.
Then, a confidence-based voting mechanism is used to combine the conclusions, i.e., $\overline{\bC}=\sum_{k=1}^3 c_k{\bC_{k}}$,
where $\overline{\bC}\in\mathbb{R}^{H\times W}$ denotes the cross-integrated conclusion map and $c_k\geq0$ is the confidence level.
In general, one can determine $c_k$ by evaluating the strength of each fuzzy rule.
In this study, we employ the equal confidence, $c_k=\frac{1}{3}$, to integrate these fuzzy conclusions.
Second, to obtain the final integrated conclusion $\bC\in\mathbb{R}^{H\times W}$, an index-based combination is employed to combine the crisp conclusion (i.e., $\widetilde{\bM_{\textbf{R4}}}$ and $\widetilde{\bM_{\textbf{R5}}}$) and $\overline{\bC}$, 
\begin{align}\label{eq: fuzzy_conclusion}
    [\bC]_{i,j} = 
    \begin{cases}
        1, & \text{if } {(i,j)} \in \Omega_{\text{max}}^{e1} \cap \Omega_{A}, \\
        0, & \text{if } {(i,j)} \in \Omega_{\text{min}}^{e2} \cap \Omega_{B}, \\
        [\overline{\bC}]_{i,j}, & \text{otherwise}.
    \end{cases}
\end{align}
where $\Omega_{A}$ and $\Omega_{B}$ denote the sets of indices at which elements within $\widetilde{\bM_{\textbf{R4}}}$ and $\widetilde{\bM_{\textbf{R5}}}$ are classified into 1, respectively.
Besides, $\Omega_{\text{max}}^{e1}$ and $\Omega_{\text{min}}^{e2}$ are the sets of indices corresponding to  the top $e1$ largest and $e2$ smallest values of $\overline{\bC}$, respectively.
In this study, we empirically set $e1:=20\%$ of $|\Omega_{A}|$ and $e2:=20\%$ of $|\Omega_{B}|$.
Based on the parameter analysis presented in Supplementary Figure 1, the proposed classical MCDM framework demonstrates robustness across varying parameter settings.
Overall, the averaged AUC scores across the eight real-world HAD datasets (see Figure \ref{fig:data}) suggest relatively small $e1$ and $e_2$, approximately $10\%$ to $20\%$ of $|\Omega_{A}|$ and $|\Omega_{B}|$, respectively.
These observations closely match the default settings of $e1:=20\%$ and $e_2:=20\%$.
Thus far, we have completed the implementation of the FC step.

Eventually, the defuzzification step (see Figure \ref{fig:classical_Fuzzy_detection}) is designed to drive a more certain outcome (i.e., CFD) from the integrated conclusion $\bC$.
In the first step of the defuzzification, we aim to surpass the energy of background scores while further enhancing the intensity of the anomaly, i.e., energy-contrastive (EC) enhancement. 
To this end, we apply a nonlinear EC transform on $\bC$, resulting in the EC map $\widetilde{\bD_C}$ with $[\widetilde{\bD_C}]_{i,j}=1-\exp(-\alpha[\bC]_{i,j})$, where $\alpha\geq0$ is the EC factor.
In this study, $\alpha$ is automatically determined by
\begin{align}\label{energy_balance}
    \alpha^{\star}=\arg \min_{\alpha\geq0}~ \frac{1}{2}\sum_{(i,j)} \left( 1-e^{(-\alpha[\bC]_{i,j})}-[\bF_{\text{max}}]_{i,j} \right)^2,
\end{align}
where $\bF_{\text{max}}\in\mathbb{R}^{H\times W}$ denotes the fuzzy degree map (i.e., $\bF_M$, $\bF_G$, and $\bF_S$) containing highest energy. 
Since the objective function in \eqref{energy_balance}, denoted as $\mathcal{L}(\alpha)$, is nonconvex and contains transcendental functions, strictly pursuing the optimal solution may lead to less efficient inference.
Thus, we solve it using the gradient descent method (GDM) \cite{gradient_descent} without the non-negative constraint, leading to its gradient,
\[
\frac{d\mathcal{L}(\alpha)}{d\alpha} = \sum_{(i,j)} \left( 1 - e^{-\alpha[\bC]_{i,j}} - {[\bF_{\text{max}}]_{i,j}} \right) [\bC]_{i,j} e^{-\alpha [\bC]_{i,j}},
\]
followed by projecting the solution back to the non-negative orthant. 
Here, we initialize $\alpha:=1$, learning rate of 2, and the maximum iteration of 1E4 for the GDM.
Once we have $\alpha$, we leverage the guided filter \cite{Guided_filter} to preserve the spatial smoothness, and the CFD can eventually be obtained by $\bD_C=\text{Guided Filter}(\widetilde{\bD_C})$.
\subsection{Quantum Multi-Criteria Decision Making}\label{subsec: QFD}
\begin{figure*}[t]
    \centering
    \includegraphics[width=1\linewidth]{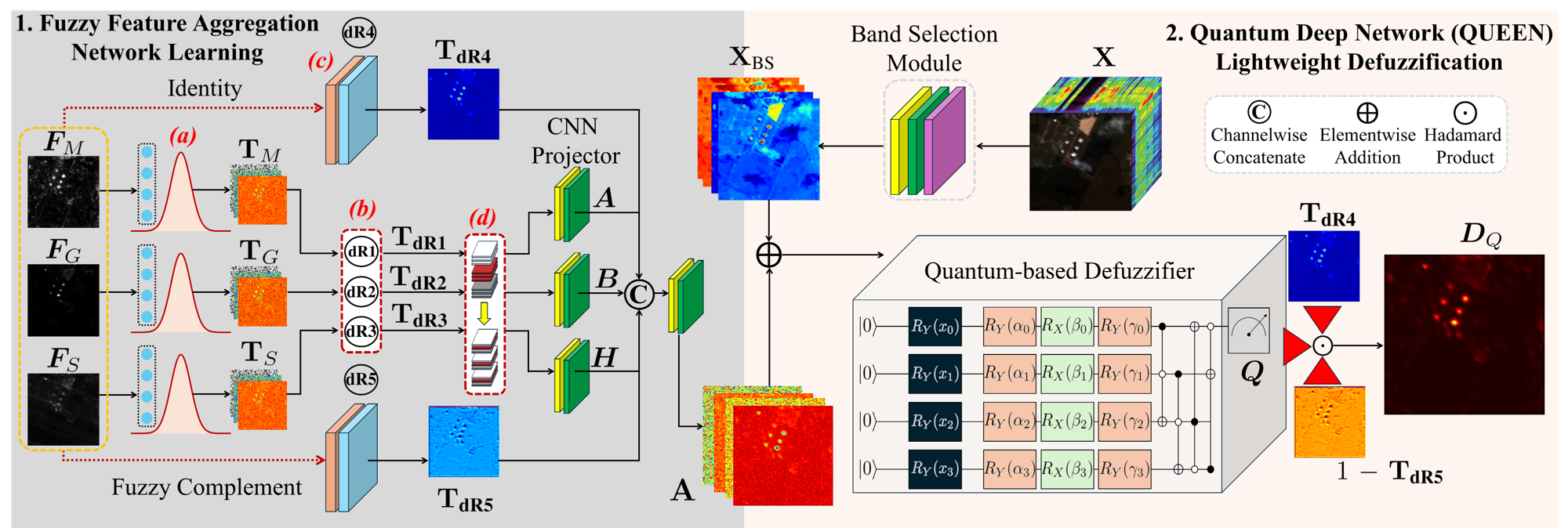}
    \caption{Overall architecture of the proposed quantum MCDM. 
    Among the flowchart, the yellow, purple, green, and blue blocks denote a full convolutional layer, a depth-wise convolutional layer, an activation function, and the soft rounding operator, respectively.
    The red markers labeled in \textit{(a)} to \textit{(d)} correspond to the modules illustrated in Figure \ref{fig:Fuzzy_detection_modules}.
    Further implementation details are presented in Section \ref{subsec: QFD}.}\label{fig:Fuzzy_detection}
    \vspace{-0.3cm}
\end{figure*}

\begin{figure}[t]
    \centering
    \includegraphics[width=1\linewidth]{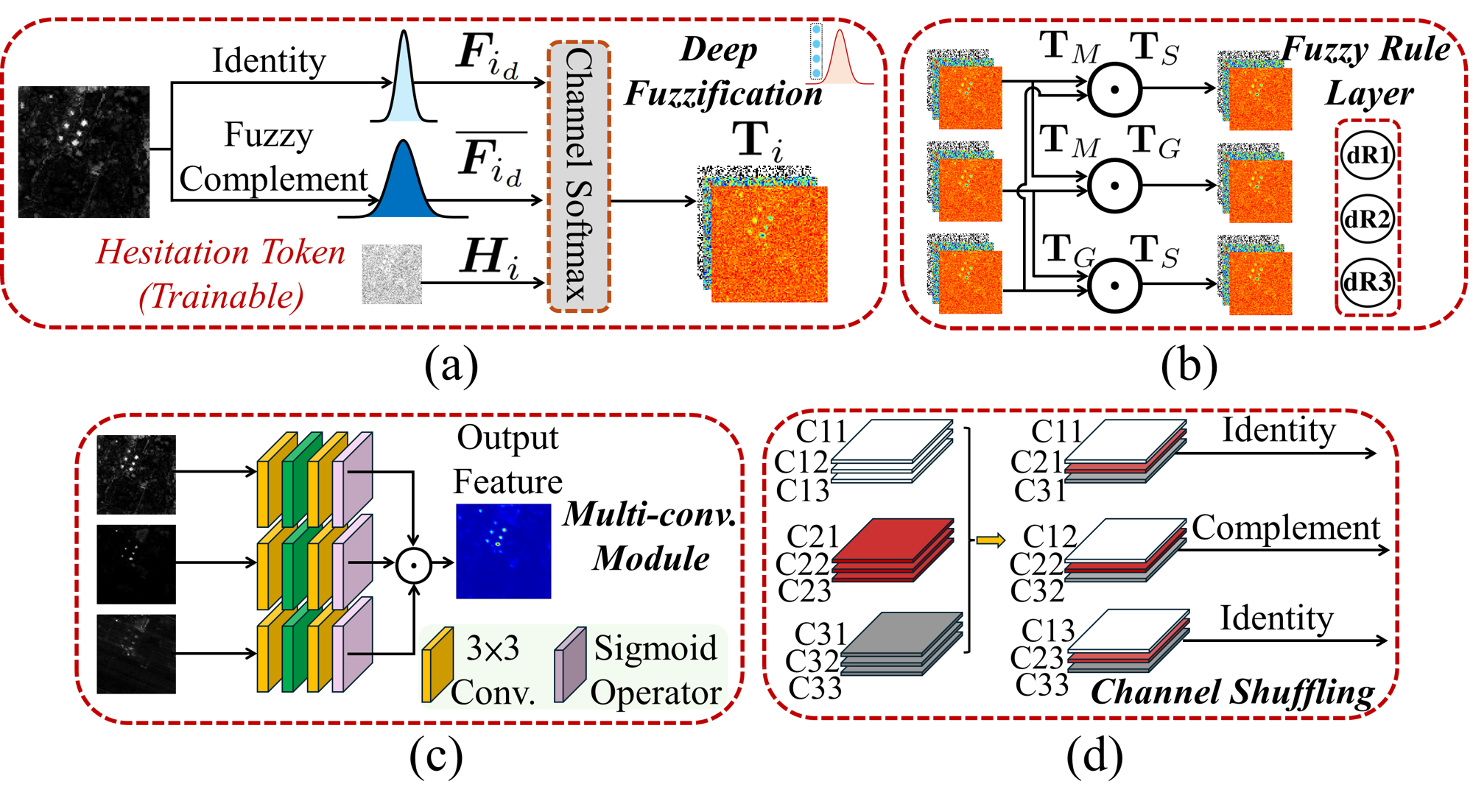}
     \vspace{-0.5cm}
    \caption{Detailed architectures of (a) deep fuzzification, (b) fuzzy rule layer, (c) multi-convolutional module, and (d) channel shuffling within the quantum MCDM.}\label{fig:Fuzzy_detection_modules}
    \vspace{-0.3cm}
\end{figure}
%
\begin{figure}[t]
    \centering
    \includegraphics[width=1\linewidth]{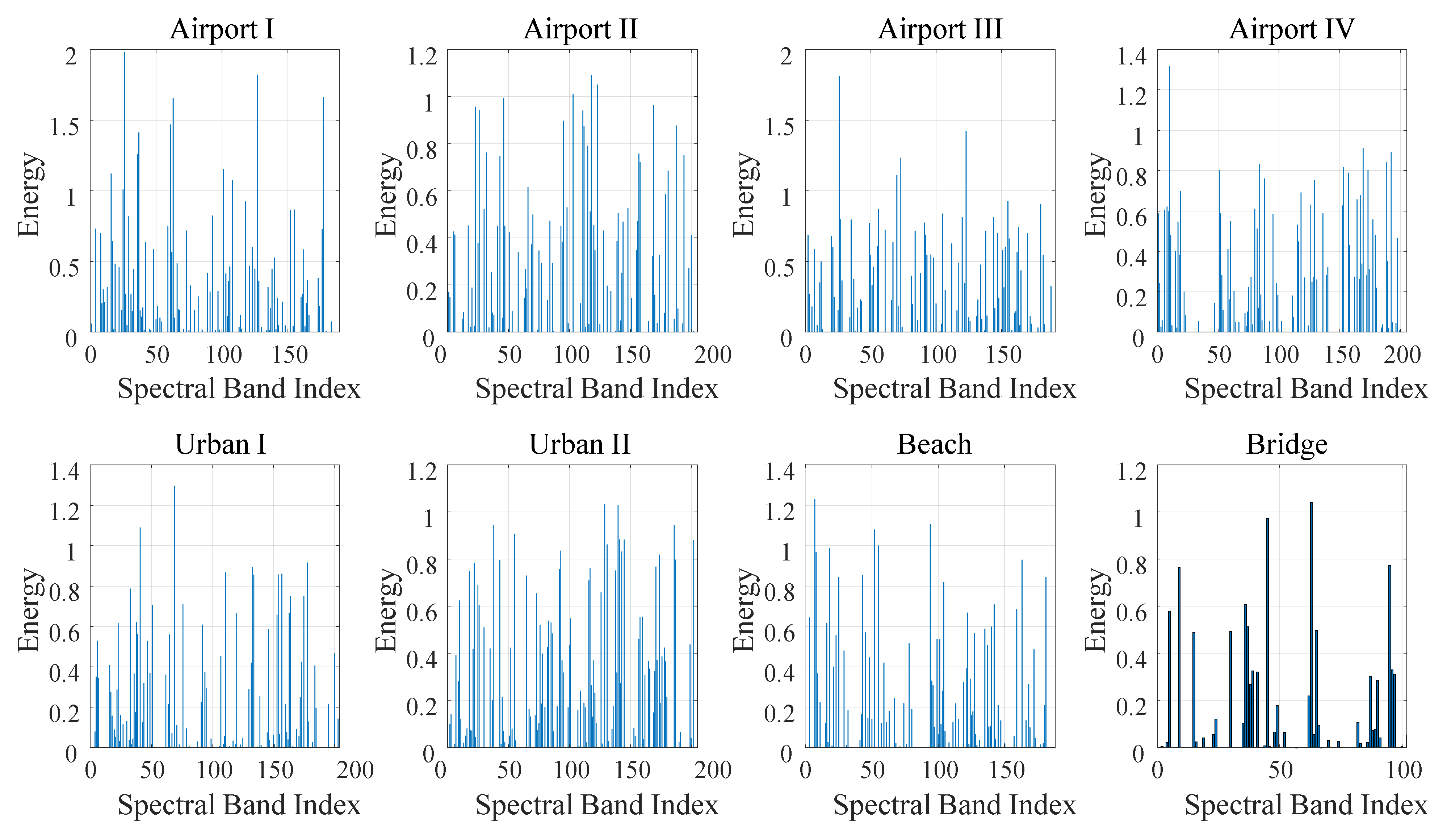}
    \caption{Energy distributions of the band-selected HSI obtained via the band-selection module (BSM).
    As defined in \eqref{EQ: BSM}, the depthwise convolutional layer first assigns a weight to each input spectral band, and a ReLU function subsequently suppresses the negatively weighted bands, thereby enabling automatic band-selection. 
    Across the eight real-world datasets (see Figure \ref{fig:data}), approximately 50\% of the spectral bands are automatically identified as redundant and hence are suppressed, resulting in the observed null-energy bands.
    Further analysis regarding the band selection results is presented in Supplementary Figure 2.
    }\label{fig:BS_result}
    \vspace{-0.3cm}
\end{figure}

In this section, we elaborate on the implementations of the proposed quantum MCDM (see Figure \ref{fig:Fuzzy_detection}), including the fuzzy feature aggregation network, band selection module (BSM), and quantum-based defuzzifier.
First, in the fuzzy feature aggregation network, we further extend the five fuzzy rules [see \eqref{equ: five_fuzzy_rule} and \eqref{equ: five_fuzzy_rule2}] into deep fuzzy rule-based tracks (denoted as \textbf{dR1}-\textbf{dR5}) to capture the interpretable fuzzy features.
Specifically, in the \textbf{dR1}-\textbf{dR3} tracks, we leverage the trainable Gaussian MFs to further fuzzify $\bF_M$, $\bF_G$, and $\bF_S$ (also their complements), resulting in the deep fuzzy representations (DFRs), i.e., 
\begin{align}
\bF_{i_d}&=G_{\theta_{ai}}(\bF_{i}),~\forall i\in\{M,G,S\},\notag
\\
\overline{\bF_{i_d}}&=G_{\theta_{bi}}(1-\bF_{i}),~\forall i\in\{M,G,S\},
\end{align}
where $\bF_{i_d}$ and $\overline{\bF_{i_d}}\in\mathbb{R}^{H\times W}$ denote the DFR of $\bF_i$ and $\bF_i^c$, respectively, under the specific HAD property (i.e., morphological $M$, geometrical $G$, and statistical $S$).
Moreover, $G_{\theta}(\cdot)$ denotes the trainable Gaussian MF with parameters $\theta$.
In this manner, $\bF_{i_d}$ and $\overline{\bF_{i_d}}$ can be interpreted as the degree of anomaly and background, respectively, in a specific deep fuzzy space.
To further upgrade the flexibility for the fuzzy feature learning, we leverage trainable hesitancy tokens, $\bH_i,~\forall i\in\{M,G,S\}$, to denote the degree of hesitancy.
In detail, given a degree of anomaly $a$ and degree of background $b$, the concept of extended fuzzy set \cite{extenedFSs} allows the pixel not only to be an anomaly or background (i.e., $a+b=1$) but also to contain a degree of hesitancy $h$, i.e., [$h=1-(a+b)$], leading to a higher flexibility and robustness (to be detailed in Section \ref{subsec: Qualitative and Quantitative Analysis}).
In the proposed quantum MCDM, hesitancy tokens are initialized as all-zero tensors to ensure a better performance and robustness.
To provide further insights, we evaluate the detection performance of the quantum MCDM with different initialization schemes (for hesitancy tokens), including all-one, normal distribution, uniform distribution, and all-zero initialization.   
As presented in Supplementary Table I, the random initializations from normal and uniform distributions achieve comparable yet promising performance, indicating that the hesitancy tokens do not rely on a specific initialization strategy.
In addition, we observe that all-one and all-zero initialization yield significantly different results (see Supplementary Table I).
Since each input fuzzy degree map, $\bF_M$, $\bF_G$, and $\bF_S$, is capable of presenting certain types of anomalies preliminarily (see Table \ref{tab: ablation study MF}), the hesitancy degrees are relatively low; therefore, initializing them to excessively large values (e.g., all-one initialization) may be less suitable.
In contrast, when initialized them to all-zero tensors, the hesitancy degrees can be gradually adjusted during optimization, thereby enabling the proposed framework to handle challenging real-world scenarios.
Consequently, this initialization yields effective detection (see Supplementary Table I) and is therefore employed in the proposed quantum MCDM framework.
To satisfy the sum-to-one property, we stack the anomaly, background, and hesitancy maps along the channel direction, followed by a softmax operator, i.e.,
\begin{align}
    \mathbf{T}_i=\text{Softmax}\left(\text{Cat}(\bF_{i_d},\overline{\bF_{i_d}},\bH_i)\right), \forall i\in\{M,G,S\}, 
\end{align}
where $\mathbf{T}_i\in\mathbb{R}^{H\times W \times 3}$ and $\text{Cat}(\cdot)$ denotes the stacked feature tensor under the specific HAD property (i.e., $M$, $G$, and $S$) and the concatenate operator, respectively.
The above deep fuzzification step refers to the red marker \textit{(a)} in Figure \ref{fig:Fuzzy_detection}, where the corresponding details are graphically presented in Figure \ref{fig:Fuzzy_detection_modules}(a).
Then, in the fuzzy rule layer [see the red marker \textit{(b)} in Figure \ref{fig:Fuzzy_detection}], we adapt the algebraic product, rather than the Einstein operators used in the classical MCDM, to efficiently match the degree for $\textbf{dR1}$-$\textbf{dR3}$ tracks in the quantum MCDM,
\begin{align}
\mathbf{T}_{\textbf{dR1}}&=\mathbf{T}_M\odot\mathbf{T}_S,\notag
\\
\mathbf{T}_{\textbf{dR2}}&=\mathbf{T}_M\odot\mathbf{T}_G,\notag
\\
\mathbf{T}_{\textbf{dR3}}&=\mathbf{T}_G\odot\mathbf{T}_S,
\end{align}
where $\mathbf{T}_{\textbf{dR}k}\in\mathbb{R}^{H\times W\times 3},~\forall k \in \mathcal{I}_3$ denotes the deep matching degree tensor (DMDT) of the $k$th track.
Specifically, in the classical MCDM, the fuzzy degree maps are used (without learnable processing) to conduct MCDM for real-time computing.
To derive additional information from these fixed degrees, fuzzy operators must provide gradual degree transitions (e.g., Einstein operators) for effective performance (see Section \ref{subsubsec: ablation_MCDM}).
In contrast, the learnable design of quantum MCDM already offers strong flexibility for modeling complex scenarios, thereby effectively enhancing detection performance of the proposed HyFuHAD (see Section \ref{subsubsec: ablation_MCDM}).
Accordingly, the fuzzy operators in the quantum MCDM are expected to be simpler for ensuring the smoothness of the optimization process, rather than providing high flexibility.
In light of this perspective, the more straightforward algebraic product is therefore employed in the quantum MCDM.
As for $\textbf{dR4}$ and $\textbf{dR5}$ tracks, we employ the shared weights multi-convolutional modules as deep MFs [see the red marker \textit{(c)} in Figure \ref{fig:Fuzzy_detection}] for deep fuzzification, followed by performing the fuzzy conjunction to match the rules, and soft rounding $\text{R}_s(\cdot)$ to increase the certainty, 
\begin{align*}
\mathbf{T}_{\textbf{dR4}}&=\!\text{R}_s\left(C_{\theta_{R4}}(\bF_M)\odot C_{\theta_{R4}}(\bF_G)\odot C_{\theta_{R4}}(\bF_S)\right),\notag
\\
\mathbf{T}_{\textbf{dR5}}&=\!\text{R}_s\left(C_{\theta_{R5}}(1\!-\!\bF_M)\odot C_{\theta_{R5}}(1\!-\!\bF_G)\odot C_{\theta_{R5}}(1\!-\!\bF_S)\right),
\end{align*}
where $\mathbf{T}_{\textbf{dR4}}$, $\mathbf{T}_{\textbf{dR5}}\in\mathbb{R}^{H\times W}$ are the DMDT of $\textbf{dR4}$ and $\textbf{dR5}$ track, respectively.
Besides, $C_{\theta}(\cdot)$ is the function of the convolutional MF with parameter $\theta$. 
Subsequently, to integrate the DMDTs into a single discriminative representation, we first rearrange the features in $\mathbf{T}_{\textbf{dR1}}$, $\mathbf{T}_{\textbf{dR2}}$, and $\mathbf{T}_{\textbf{dR3}}$ by the channel shuffling [see the red marker \textit{(d)} in Figure \ref{fig:Fuzzy_detection}].
Specifically, the first, second, and third features of these DMDTs correspond to the degree of anomaly, background, and hesitancy, respectively, under the different rules.
Therefore, features of the same type are rearranged into a unified tensor, then applied to a convolutional projector for the feature aggregation, i.e.,  
\begin{align}
\bA &=f_\text{Proj}^1\left(\text{Cat}([\mathbf{T}_{\textbf{dR1}}]_{:,:,1},[\mathbf{T}_{\textbf{dR2}}]_{:,:,1},[\mathbf{T}_{\textbf{dR3}}]_{:,:,1})\right),\notag
\\
\bB &=f_\text{Proj}^1\left(1-\text{Cat}([\mathbf{T}_{\textbf{dR1}}]_{:,:,2},[\mathbf{T}_{\textbf{dR2}}]_{:,:,2},[\mathbf{T}_{\textbf{dR3}}]_{:,:,2})\right),\notag
\\
\bH &=f_\text{Proj}^1\left(\text{Cat}([\mathbf{T}_{\textbf{dR1}}]_{:,:,3},[\mathbf{T}_{\textbf{dR2}}]_{:,:,3},[\mathbf{T}_{\textbf{dR3}}]_{:,:,3})\right),
\end{align}
where $\bA$, $\bB$, and $\bH\in\mathbb{R}^{H\times W}$ denote the aggregated anomaly, background, and hesitancy feature, respectively. 
Moreover, $f_\text{Proj}^1(\cdot)$ represents the CNN projector composed of a $1\times 1$ convolution layer and Leaky ReLU with a negative slope of 0.2 (see Figure \ref{fig:Fuzzy_detection}).
Eventually, the discriminative fuzzy feature $\mathbf{A}\in\mathbb{R}^{H\times W \times D}$ can be simply obtained by $\mathbf{A} =f_\text{Proj}^1\left(\text{Cat}(\bA,\bB,\bH,\mathbf{T}_{\textbf{dR4}},\mathbf{T}_{\textbf{dR5}})\right)$, where $D:=4$ is the given dimension of deep space. 

Consequently, we employ the quantum-based defuzzifier to infer the QFD from the fuzzy feature and HSI (see Figure \ref{fig:Fuzzy_detection}).
To achieve an efficient inference, we further leverage the BSM, comprising a depthwise convolutional layer $f_\text{Depth}(\cdot)$, ReLU, and a regular convolutional layer $f_\text{CNN}(\cdot)$, to select the $D:=4$ discriminative bands, i.e.,
\begin{align}\label{EQ: BSM}
   \mathbf{X}_{\text{BS}}=f_\text{CNN}(\text{ReLU}(f_\text{Depth}(\mathbf{X})))\in\mathbb{R}^{H\times W \times D}. 
\end{align}
All convolutional layers in BSM employ a kernel size of $1\times 1$.
To further enhance the understanding of such automatic band-selection, the corresponding energy distributions of band-selected HSIs are presented in Figure \ref{fig:BS_result}. 
Specifically, in BSM, the depthwise convolutional layer first assigns a trainable weight to each input spectral band, followed by a ReLU function that suppresses the ones associated with negative weights to enable the automatic band selection procedure.
Therefore, the averaged energy level of each band reflects the selection results, wherein the null-energy bands indicate redundant information identified by BSM.
As illustrated in Figure \ref{fig:BS_result}, in each adjacent spectral wavelength, approximately 50\% of the spectral bands are automatically suppressed to mitigate the spectral redundancy \cite{10410870}. 
To validate the effectiveness of the band selection mechanism, the naive HAD method GRX \cite{RX}, which identifies anomalies solely based on input structure, is employed for further evaluations.
When the full-band HSIs are used as input, the averaged AUC score across the eight real-world datasets reaches only 0.95, while it significantly improves to 0.97 when the band-selected HSIs are adopted.
This advantage may be attributed to the ability of BSM to yield a subset of bands with strong discriminative capability. 
 As illustrated in Supplementary Figure 2, the selected bands can form highly distinguishable signatures between the anomaly and background components.
Furthermore, the unselected ones include noisy, uninformative, and corrupted bands, which indeed have only a limited or even negative contribution to the HAD task.
With such an effective design, the proposed quantum MCDM requires a single regular convolutional layer to fuse the informative bands, rather than incorporating overly complex architectures for feature extractions.
Next, we present the design of the quantum-based defuzzifier and emphasize its advantages as follows.
First, it is worth noting that advanced quantum technologies (e.g., QUEEN) have been applied across diverse applications. 
For instance, QUEENs have demonstrated their superiority in the RS technologies, such as restoration \cite{HyperQUEEN,HyperKING}, multispectral unmixing \cite{PRIME}, change detection \cite{QUEENG}, and mangrove monitoring \cite{QEDNet}.
Due to their unique unitary-computing mechanism, QUEENs facilitate encoding of input features into the complex-valued quantum space \cite{CYGNSS_arXiv,Drought_arXiv}, followed by using the unitary quantum gates \cite[Table I]{HyperQUEEN} to learn target quantum states.
Beyond the RS applications, QUEENs are also adopted as a complex MF \cite{TFS_quantum_classification} and defuzzifier \cite{Quantum_DEFUZZIFIER} for advanced classification tasks in the fuzzy systems community.
The high flexibility and unique information of QUEENs directly contribute significant performance upgrades for fuzzy systems.
These advantages collectively motivate us to employ the QUEEN (with a sigmoid function as the last layer) as a defuzzifier.

In the proposed framework, the QUEEN is developed based on both theoretical groundings and practical considerations.  
First, increasing the depth of quantum circuits may increase the risk of the Barren Plateaus (BP) effect (i.e., gradient vanishing issue of QUEENs).
Specifically, the number of layers $n$ is generally scaled with the number of qubits $k$ to ensure sufficient expressibility for processing information-dense inputs \cite{sim2019expressibility}.
However, as theoretically proven in \cite[Theorem 1]{HyperQUEEN}, the expectation and variance of the gradient in a QUEEN approach zero as $k\rightarrow\infty$.
In other words, QUEENs with overly deep architectures become difficult to train using gradient-based optimizers. 
Accordingly, we employ the full expressibility (FE) \cite{HyperQUEEN} architecture, ``$R_Y-R_X-R_Y$'' (see Figure \ref{fig:Fuzzy_detection}) to process the encoded quantum states.
The FE property enables QUEENs to realize any valid unitary operators within a single quantum circuit.
This capability indicates that three rotation layers appear to be sufficient for effective performance, without using excessively complex architectures as in classical deep learning.
Moreover, due to the FE property, the $R_Y-R_X-R_Y$ layer is expected to generalize across all fuzzy features; therefore, we adopt a share-weighted strategy in our implementation (see Supplementary Figure 3).
Consequently, even when the spatial dimension of input HSI increases, the proposed framework remains deployable on a 4-physical-qubit quantum device.

Second, we elaborate on the design of the shallow Toffoli entanglement layer (see Supplementary Figure 3).
Since the additional readout error of each physical qubit is relatively non-negligible (about 2-3\%) \cite{arute2019quantum}, we measure only a single qubit and preserve the quantum information via a full entanglement (i.e., direct entanglement among all physical qubits).
To realize the design, the Toffoli entanglement layer can be implemented using either the three-qubit “CCNOT” gate or the two-qubit “CNOT” gate \cite[Table I]{HyperQUEEN}.
The former requires a depth of 4 to achieve full entanglement, whereas the latter stacks six quantum gates (i.e., qubits 1-2, 1-3, 1-4, 2-3, 2-4, and 3-4) for this aim.
The depth selection substantially affects its practical applicability on near-term quantum hardware. 

In an ideal setting, quantum
gates should be able to implement the corresponding target unitary matrix \cite[Table I]{HyperQUEEN}, yielding a 100\% fidelity. 
However, practically, quantum gates exhibit imperfect fidelity due to noise effects, indicating the presence of approximation error within the quantum circuit.
For example, single-qubit gates typically achieve 99.9\% fidelity \cite{bluvstein2024logical}, whereas the fidelity decreases to approximately 98\% for multi-qubit gates \cite{evered2023high}. 
Therefore, the overall fidelity decreases exponentially with circuit depth.  
Moreover, quantum computers require maintaining quantum coherence throughout executions \cite{zurek1991decoherence}.
However, the coherence time of near-term quantum devices remains relatively limited.
Thus, the feasible depth of a quantum circuit can be
approximated as the coherence time divided by the gate inference time \cite[Section 8]{nielsen2001quantum}.
Accordingly, both quantum coherence and gate fidelity again suggest a shallow architecture, which motivates us to adopt the CCNOT Toffoli entanglement layer.

With the above designs, the output quantum feature $\bQ\in\mathbb{R}^{H\times W}$ can be obtained by
\begin{align}
    \bQ=f_{\text{QUEEN}}(\mathbf{A}+\mathbf{X}_{\text{BS}}),
\end{align}
where $f_{\text{QUEEN}}(\cdot)$ denotes the entire function of the quantum-based defuzzifier (see Supplementary Figure 3 for the inference mechanism).
Eventually, we can have the QFD $\bD_Q=\bQ\odot\mathbf{T}_{\textbf{dR4}}\odot (1-\mathbf{T}_{\textbf{dR5}})$.
As for the loss function design, we integrate the concepts of semi-supervised learning \cite{CODEHCD} and pseudo-labeling \cite{pham2021meta}.
Specifically, we adopt the elements of the binarized CFD [i.e., $\overline{\bD_C}=\text{K-means}(\bD_C)]$ that correspond to a high possibility of being an anomaly or background as the pseudo labels.
Furthermore, we employ the total variation (TV) regularization \cite{tvpytorch} to promote the spatial smoothness of $\bD_Q$, resulting in the total loss function, i.e.,
\begin{align*}
    \mathcal{L}(\bD_Q,\!\overline{\bD_C}):=\!\frac{1}{|\Omega_{\text{p}}|} \!\!\sum_{(i,j)\in \Omega_{\text{p}}}\!\!\!\text{BCE}([\bD_Q]_{i,j},[\overline{\bD_C}]_{i,j})\!+\!\lambda\text{TV}(\bD_Q),
\end{align*}
where $\Omega_{\text{p}}=\Omega_{\text{max}}^{e3}\cup\Omega_{\text{min}}^{e4}$, with $e3$ and $e4$ denotes 10\% of anomaly pixels (i.e., number of 1's) and 10\% of background pixels (i.e., number of 0's), respectively, in $\overline{\bD_C}$.
Besides, $\text{BCE}(\cdot)$ and $\lambda:=$5E-5 denote the binary cross-entropy \cite{QEDNet} and trade-off parameter, respectively.
Accordingly, we have completed the implementation of the proposed QFD.

\section{Experimental Results}\label{sec: Experiment}
In this Section, we conduct comprehensive experiments and analyses to substantiate the superiority of the proposed HyFuHAD.
The details of diverse real-world datasets and experimental settings are summarized in Section \ref{subsec: Datasets and Settings}.
Next, qualitative and quantitative analyses are organized in Section \ref{subsec: Qualitative and Quantitative Analysis}.
Besides, we provide the ablation study and discussions in Section \ref{subsec: ablation}. 
\subsection{Datasets and Settings}\label{subsec: Datasets and Settings}
%
\begin{table}[t]
\centering
\caption{Data information of the representative real-world scenarios presented in Figure \ref{fig:data}, with the number of abnormal pixels $\underline{A}$, and the image size of $W\times W$ pixels.}\label{tab: data description}
\begin{tabular}{cccccc} \hline\hline
Datasets  & Location  & $W$  & $C$ & Resolution  & $\underline{A}$ 
\\ \hline
Airport \uppercase\expandafter{\romannumeral 1}&  San Diego &  $90$   & 189 & 3.5 m & 77
\\
Airport \uppercase\expandafter{\romannumeral 2} & Los Angeles & $100$    & 205 & 7.1 m  & 87
\\
Airport \uppercase\expandafter{\romannumeral 3} & Gulfport & $100$   & 191 & 3.4 m  & 60
\\
Airport \uppercase\expandafter{\romannumeral 4} & Los Angeles & $100$     & 205 & 17.2 m & 170
\\
Urban \uppercase\expandafter{\romannumeral 1} & Texas Coast &   $100$   & 204  & 7.1 m  & 67
\\
Urban \uppercase\expandafter{\romannumeral 2}& San Diego &   $100$   & 205  & 3.4 m  & 272
\\
Beach   & Bay Champagne &   $100$   & 188  & 4.4 m  & 11
\\
Bridge   & Pavia City &   $100$   & 102  & 1.3 m  & 68
\\
\hline\hline
\end{tabular}
\end{table}
%
\begin{figure}[t]
    \centering
    \includegraphics[width=1\linewidth]{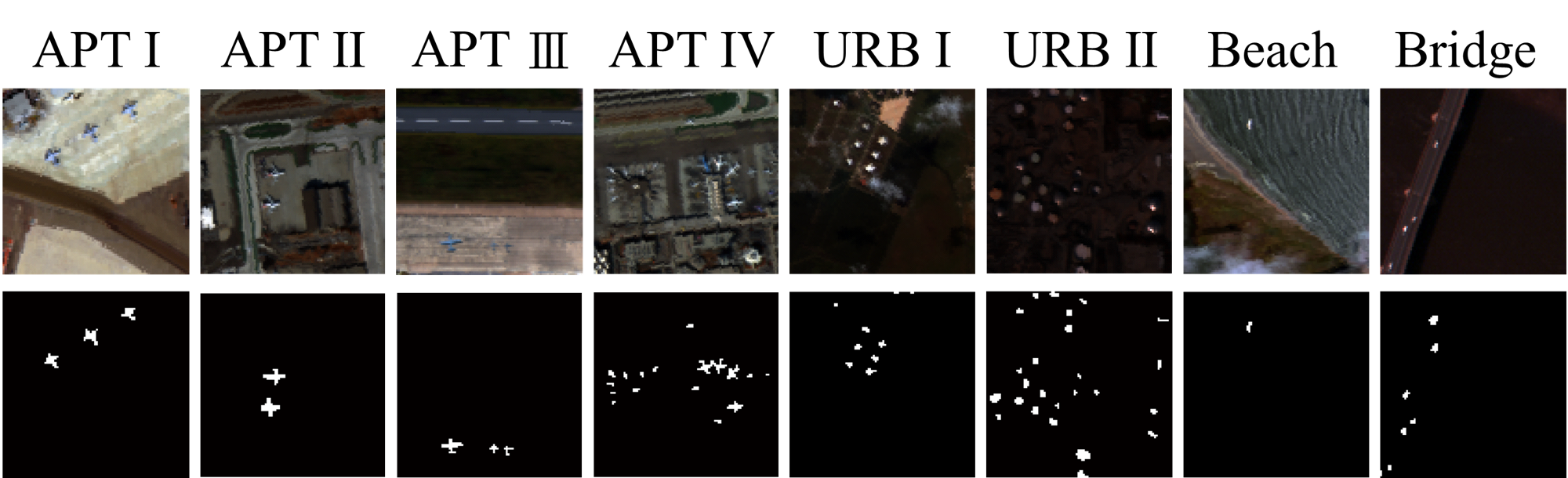}
    \caption{False-color compositions of the eight representative real-world datasets used in our experiments, where the corresponding reference maps and the qualitative assessments are presented in Figure \ref{fig:detection_comparison}.}\label{fig:data}
    \vspace{-0.3cm}
\end{figure}
%
To comprehensively evaluate the proposed HyFuHAD, we employ the diverse and representative real-world datasets \cite{SuperRPCA,TGFA-AD,TAEF,GT-HAD}, as graphically presented in Figure \ref{fig:data}.
Among them, the Bridge dataset is captured by the Reflective Optics System Imaging Spectrometer (ROSIS) sensor \cite{TGFA-AD}, while the remainder are acquired by NASA's Airborne Visible/Infrared Imaging Spectrometer (AVIRIS) sensor \cite{AVIRISdata}.
More characteristics of these datasets are summarized in the Table \ref{tab: data description}.

Diverse types of benchmarks (i.e., statistical, optimization-based, as well as SOTA DL-based approaches), including GRX \cite{RX}, TPCA \cite{TPCA}, GTVLRR \cite{GTVLRR}, Auto-AD \cite{Auto-AD}, LARTVAD \cite{LARTVAD}, BockNet \cite{BockNet}, TAEF \cite{TAEF}, PUNNet \cite{PUNNet}, GTHAD \cite{GT-HAD}, TGFA-AD \cite{TGFA-AD}, DirectNet \cite{DirectNet}, NL2Net \cite{NL2Net}, BSDM \cite{BSDM}, and OT-AD \cite{OT-AD} are incorporated for a persuasive evaluation.
Nearly all default settings provided by their official implementations are adopted throughout the experiments.
Due to space limitations, we summarize their experimental settings in Supplementary Note 1.
For the proposed FuHAD, the trainable Gaussian MFs and quantum defuzzifier are initialized from a Gaussian distribution.
Moreover, the convolutional layers in the quantum MCDM framework are initialized using the Kaiming-uniform scheme (default setting in the PyTorch library).
The initial learning rate of the Adam optimizer \cite{kingma2014adam} is set to 3.5E-3.
Regarding the number of training iterations, we observe that the minimum epochs to achieve generalizable and effective detection is around 30.  
Accordingly, the iteration number is empirically set to 30 to ensure computational efficiency.  
The computational equipment, equipped with an NVIDIA RTX 3090 GPU and an AMD Ryzen 5950X CPU with a 3.40-GHz speed (128 GB of memory), is used to conduct all experiments under the Ubuntu 22.04.5 LTS system.
The software environments include Pennylane 0.34.0 (for quantum simulation), Python 3.10.9, PyTorch 2.0.0, and MATLAB R2023a.
\subsection{Qualitative and Quantitative Analysis}\label{subsec: Qualitative and Quantitative Analysis}
%
\begin{figure*}[t]
    \centering
    \includegraphics[width=1\linewidth]{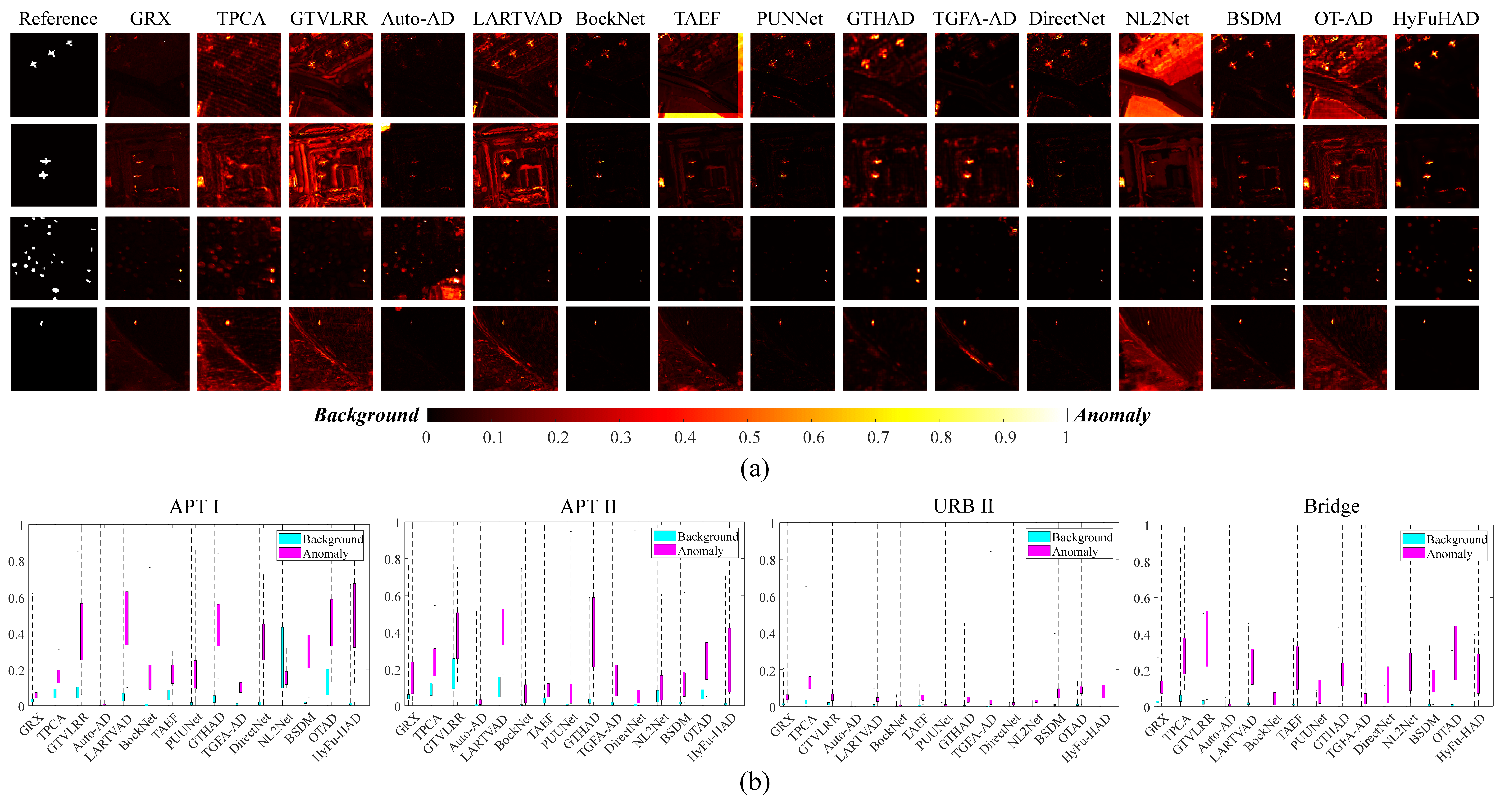}
    \caption{(a) Heatmap visualizations of the detection results and (b) Box-whisker plots of HAD algorithms across the representative real-world scenes, including Airport \uppercase\expandafter{\romannumeral 1}, Airport \uppercase\expandafter{\romannumeral 2}, Urban \uppercase\expandafter{\romannumeral 2}, and Beach datasets.
    Complete qualitative evaluations and Box-whisker plots are illustrated in Supplementary Figure 4.
    Further analytical discussions are presented in Section \ref{subsec: Qualitative and Quantitative Analysis}.
    }\label{fig:detection_comparison}
    \vspace{-0.3cm}
\end{figure*}
The qualitative evaluations across several representative scenarios are presented in Figure \ref{fig:detection_comparison}, while the complete results are collectively provided in Supplementary Figure 4 owing to space limitations. 
The qualitative analyses are presented as follows.
In the Airport \uppercase\expandafter{\romannumeral 1} and Airport \uppercase\expandafter{\romannumeral 2} scenarios [see the 1st and 2nd rows in Figure \ref{fig:detection_comparison}(a)], the statistical detector (i.e., GRX) and optimization-based approaches (i.e., TPCA, GTVLRR, and LARTVAD) fail to achieve effective detections.
Specifically, in the RX detector results, both anomaly and background scores are over-suppressed and hence indistinguishable because complex airport scenarios rarely satisfy the assumption of a multivariate normal distribution.
Moreover, optimization-based approaches that rely on specific handcrafted priors, such as TV, low-rank, and graph structure, struggle to effectively model challenging scenes. 
On the other hand, certain DL-based detectors, including Auto-AD, TAEF, PUUNet, and NL2Net, also suffer from precisely detecting anomalies, revealing the unreliability of the so-called ``black-box''. 
Although the very recent method, OT-HAD, successfully identified abnormal objects, its background suppression capability is relatively insufficient, resulting in unsatisfactory performance.
In these scenes (i.e., Airport \uppercase\expandafter{\romannumeral 1} and Airport \uppercase\expandafter{\romannumeral 2}), only BockNet, GTHAD, DirectNet, BSDM, and the proposed HyFuHAD can accurately capture the anomalies.
However, the results show that BockNet, DirectNet, and BSDM are less effective in anomaly detectability, while GTHAD suffers from insufficient background suppressibility.
In contrast to these methods, our HyFuHAD achieves both effective detectability and suppressibility on the Airport datasets.

To further demonstrate the effectiveness of the proposed HyFuHAD, the corresponding Box-whisker plots are illustrated in Figure \ref{fig:detection_comparison}(b). 
Given an estimated detection map, the detection scores can be categorized into anomaly and background classes according to the reference map.
Within each class, the top 25\% and bottom 25\% scores are discarded to more clearly and generally reflect the separability between anomaly and background scores.
The resulting score distributions are visualized as the Box-whisker plots. 
In the Airport \uppercase\expandafter{\romannumeral 1} and Airport \uppercase\expandafter{\romannumeral 2} scenarios, the proposed HyFuHAD demonstrates the highest (or second-highest) anomaly boxes, implying strong anomaly detectability.
Furthermore, HyFuHAD exhibits the lowest background boxes, even compared with the top-performing baselines (e.g., GTHAD, DirectNet, and BSDM).
These experimental results collectively validate the effectiveness of our HyFuHAD on the airport datasets.

Subsequently, we analyze the results on Urban \uppercase\expandafter{\romannumeral 2} dataset.
As observed in the third row of Figure \ref{fig:detection_comparison}(a), most benchmarks fail to highlight the anomalies, excepting TPCA, Auto-AD, BSDM, OT-AD, and the proposed HyFuHAD.
Nevertheless, the former two HAD baselines (i.e., TPCA and Auto-AD) struggle with background suppression capability.
Specifically, TPCA falsely highlights the upper-right background region as anomalies, limiting the overall qualitative performance.
In the Auto-AD result, the false-detections in the lower-right area are visually noticeable. 
By comparison, the proposed HyFuHAD, as well as the two recent approaches (i.e., BSDM and OT-AD), achieve both promising anomaly detectability and strong background suppressibility.
Nevertheless, the proposed HyFuHAD still outperforms both BSDM and OT-AD, as substantiated by the Box-whisker plots [see Figure \ref{fig:detection_comparison}(b)].
To be more specific,  compared with these baselines, HyFuHAD exhibits higher anomaly scores with a much lower background box (barely observable) on the urban dataset. 
These results suggest that the proposed HyFuHAD achieves the strongest score separability and is therefore expected to achieve the best quantitative performance (to be detailed in the sequel).
From the above discussion, the proposed HyFuHAD achieves the best qualitative performance on the urban dataset.

In the Beach datasets [see the fourth rows of Figure \ref{fig:detection_comparison}(a)], TPCA, GTVLRR, TAEF, NL2Net, and OT-AD are less effective in suppressing the background.
On the other hand, even when compared to relatively effective benchmarks in these scenarios (e.g., GT-HAD, TGFA-AD, and DirectNet), the proposed HyFuHAD still achieves superior detections, not only due to its effective anomaly detectability but, more importantly, its superior background suppressibility. 
In the corresponding box plots, nearly all the HAD algorithms exhibit high anomaly boxes (i.e., strong detectability).
In other words, the primary differences in the qualitative performance arise from the effectiveness of background suppression.
As shown in the box plot, the background box of HyFuHAD is barely noticeable.
This observation substantially supports the strongest background suppression capability of the proposed HyFuHAD.
The remaining qualitative results are collectively presented in Supplementary Figure 4, which consistently indicate the superior qualitative performance of HyFuHAD.  
Consequently, the proposed HyFuHAD yields the best qualitative performance across the Airport, Urban, and Beach datasets; the quantitative evaluations are subsequently conducted.

To fairly substantiate the effectiveness of the proposed HyFuHAD, we further employ the receiver operating characteristic (ROC) curve \cite{SuperRPCA} and the computational time for quantitative evaluations.
In the ROC curve, the X-axis and Y-axis denote the probability of false alarm (PF) and probability of detection (PD), respectively, over the ascent-ordered thresholds $\tau\in[0,1]$.
Accordingly, when a detector identifies anomalies with high intensities while suppressing background effectively, the PD increases more rapidly than the PF, resulting in a larger area under the curve (AUC), i.e., AUC(PD, PF) score.
The overall quantitative evaluations, including the AUC scores and computational times, across the eight datasets are summarized in Table \ref{tab: performance}.
With the aforementioned in mind, we are now ready for the quantitative analysis.

Initially, as observed from Table \ref{tab: performance}, the proposed HyFuHAD consistently achieves the best quantitative performance over nearly all datasets, resulting in a superior average performance.
More importantly, the computational times of the proposed HyFuHAD are an order faster than the DL and optimization-based approaches (e.g., GTVLRR, LARTVAD, BockNet, TAEF, PUNNet, and DirectNet), substantiating its best inference efficiency except for the naive benchmark, GRX.  
These collectively substantiate the SOTA performance and promising computational efficiency of the proposed HyFuHAD compared to the benchmarks.  
To further substantiate these observed improvements, we perform a statistical significance test using the paired t-test \cite{kim2015t} to compare the proposed HyFuHAD with the top-performing baselines (i.e., GTHAD and TGFA-AD).
In detail, we consider the AUC scores obtained by the proposed HyFuHAD and the baselines across the eight datasets as two paired observations.
The comparisons are then conducted using MATLAB function ``\textit{ttest}''.
The testing results reject the null hypothesis for both comparisons at a typical 5\% significance level.
These findings indicate that the quantitative improvements achieved by the proposed HyFuHAD are statistically significant rather than marginal fluctuations.

On the other hand, the quantitative performance of the proposed quantum MCDM, with and without the novel hesitancy tokens [see Figure \ref{fig:Fuzzy_detection_modules}(a)], is evaluated under varying noise levels to verify its robustness.
Specifically, in the RS field, noise corruption is frequently reported across numerous applications \cite{zhang2023hyperspectral}.
The corruption potentially eliminates the discrepancy between the anomaly and background components, thereby increasing ambiguity (i.e., hesitancy level).
Accordingly, the quantum MCDM with hesitancy tokens is expected to demonstrate strong robustness, thereby enabling the overall HyFuHAD to achieve robust detection across varying noise levels.
In this experiment, noisy HSI inputs are simulated using additive Gaussian noise, with noise levels varying from 15 dB to 40 dB in terms of signal-to-noise ratio (SNR).
Moreover, the input dimensions of the corresponding convolutional layers within quantum MCDM are carefully adjusted after removing the hesitancy tokens.
The remaining settings, such as optimization objective and learning rates, are retained the same for fair comparisons.
The quantitative evaluations across the eight real-world datasets are illustrated in Supplementary Figure 5 (due to space limitations). 
As observed in Supplementary Figure 5, the quantum MCDM equipped with hesitancy tokens is substantially robust against noise corruption.
Even under severe noise corruption (e.g., SNR:=15 dB), the quantum MCDM with hesitancy tokens achieves an AUC of around 0.982, compared with 0.985 on the original inputs.
Attributed to such innovative design, the proposed HyFuHAD  (i.e., integration of classical MCDM and quantum MCDM with hesitancy tokens) is also barely affected by severe noise corruptions.
For instance, it still achieves an AUC of around 0.99 under a SNR of 15 dB, which is competitive with SOTA performance (see Supplementary Figure 5).
By comparison, the quantum MCDM without hesitancy tokens accurately detects anomalies only under relatively high SNR levels.
These experimental results substantially verify the strong robustness attributed to the novel hesitancy tokens.
%
\begin{table*}[t]
\footnotesize
\centering
\caption{Quantitative evaluations of the HAD algorithms over the representative real-world datasets, including AUC(PD, PF)$~\!(\uparrow)$ score and inference time. The red and blue boldfaced numbers, respectively, denote the best and second AUC(PD, PF) scores; and the inference time was reported in seconds (sec.).}\label{tab: performance}
\begin{tabular}{cc|ccccccccc} \hline \hline
                Methods &   Metrics   & Airport \uppercase\expandafter{\romannumeral 1}  & Airport \uppercase\expandafter{\romannumeral 2}     & Airport \uppercase\expandafter{\romannumeral 3}     & Airport \uppercase\expandafter{\romannumeral 4}   & Urban \uppercase\expandafter{\romannumeral 1} & Urban \uppercase\expandafter{\romannumeral 2} & Beach   &  Bridge & Average        \\ \hline
\multirow{2}{*}{GRX \cite{RX}}    & AUC(PD, PF)$~\!(\uparrow)$ & 0.9106 & 0.8404 & 0.9526 & 0.9288 & 0.9907 & 0.9887 & {\bf \red 0.9999} & 0.9887 & 0.9501 %
\\
                           & Time  & {0.0805} & { 0.1062} & { 0.0713} & { 0.0744} & {0.0826} & {0.0824} & {0.0728} & {0.0384} & {0.0761 }
                           \\ \hline
\multirow{2}{*}{TPCA \cite{TPCA}}    & AUC(PD, PF)  & 0.9075 & 0.8891 & 0.9432 & 0.9297 & 0.9391 & 0.9617 & 0.9982 & 0.9905 & 0.9449
\\%
                           & Time  & 11.0490 & 15.8763 & 14.4151 & 16.2812 & 14.9365 & 15.6777 & 13.9035 & 6.0573 & 13.5246
                           \\ \hline
\multirow{2}{*}{GTVLRR \cite{GTVLRR}}    & AUC(PD, PF)  & 0.9444 & 0.8316 & 0.9544 & 0.9420 & 0.9359 & 0.9416 & 0.9885 & 0.9846 & 0.9404
\\%
                           & Time  & 67.8305 & 99.8959 & 70.3669 & 65.4172 & 82.1170 & 60.4322 & 76.4600 & 47.3146 & 71.2293
 
                           \\ \hline
                           
\multirow{2}{*}{Auto-AD \cite{Auto-AD}}  & AUC(PD, PF)  & 0.9054 & 0.8776 & 0.9164 & 0.8365 & 0.9546 & 0.9117 & 0.9866 & 0.9557 & 0.9181   
\\%
                           & Time & 13.0297 & 5.8630 & 2.8413 & 7.1957 & 3.7920 & 1.7600 & 7.5434 & 6.0267 & 6.0065  
                           \\ \hline
\multirow{2}{*}{LARTVAD \cite{LARTVAD}}    & AUC(PD, PF)  & 0.9896 & 0.9666 & {\bf \blue 0.9940} & 0.8721 & 0.9799 & 0.9796 & 0.9940 & 0.9868 & 0.9703  
\\%
                           & Time  & 14.3245 & 18.0915 & 17.0670 & 17.5637 & 17.4289 & 17.7373 & 17.1979 & 15.7840 & 16.8994     
                           \\ \hline
\multirow{2}{*}{BockNet \cite{BockNet}}    & AUC(PD, PF)  & 0.9844 & 0.9535 & 0.9841 & 0.9366 & 0.9907 & 0.9604 & 0.9992 & 0.9753 & 0.9730 
\\%
                           & Time  & 24.8246 & 30.1987 & 28.9434 & 29.4756 & 29.8440 & 29.3982 & 29.2495 & 26.2472 & 28.5227   
                           \\ \hline
\multirow{2}{*}{TAEF \cite{TAEF}}   & AUC(PD, PF)  & 0.8538 & 0.8779 & 0.9696 & 0.9272 & 0.9827 & 0.9733 & 0.9661 & 0.9924 & 0.9429   
\\%
                           & Time  & 24.3979 & 32.2277 & 31.0263 & 31.6704 & 31.1799 & 23.0228 & 29.5896 & 27.1234 & 28.7798    
                           \\ \hline
\multirow{2}{*}{PUNNet \cite{PUNNet}}  & AUC(PD, PF) & 0.9780 & 0.9151 & 0.9586 & 0.9225 & 0.9813 & 0.9781 & 0.9704 & 0.9944 & 0.9623  
\\%
                           & Time & 21.7820 & 22.1738 & 21.8248 & 21.6153 & 21.7211 & 21.6564 & 21.1126 & 20.8890 & 21.5969     
                           \\ \hline
\multirow{2}{*}{GTHAD \cite{GT-HAD}}    & AUC(PD, PF) & {\bf \blue 0.9908} & {\bf \blue 0.9857} & 0.9908 & 0.9647 & {\bf \blue 0.9970} & 0.9868 & {\bf \blue 0.9998} & {\bf \red 0.9987} & {\bf \blue 0.9893} 
\\%
                           & Time  & 8.5533 & 11.5563 & 8.9853 & 9.5118 & 9.2304 & 9.0969 & 9.5251 & 7.6705 & 9.2662   
                           \\ \hline
\multirow{2}{*}{TGFA-AD \cite{TGFA-AD}}    & AUC(PD, PF)  & 0.9718 & 0.9642 & 0.9862 & {\bf \blue 0.9667} & 0.9959 & 0.9719 & 0.9981 & 0.9954 & 0.9813   
\\%
                           & Time & 5.5238 & 6.2315 & 6.3445 & 6.3478 & 6.3551 & 6.2189 & 6.2917 & 6.4533 & 6.2208    
                           \\ \hline
\multirow{2}{*}{DirectNet \cite{DirectNet}}    & AUC(PD, PF)  & 0.9743 & 0.8697 & 0.9576 & 0.8933 & 0.9906 & 0.9930 & 0.9880 & 0.9920 & 0.9573  
\\%
                           & Time  & 3.945E3 & 4.773E3 & 5.584E3 & 4.783E3 & 4.383E3 & 4.815E3 & 3.986E3 & 2.909E3 & 4.397E3    
                           \\ \hline
\multirow{2}{*}{NL2Net \cite{NL2Net}}    & AUC(PD, PF)  & 0.3674 & 0.6989 & 0.5862 & 0.9226 & 0.9844 & {\bf \blue 0.9965} & 0.9318 & 0.9908 & 0.8098 
\\ %
                            & Time & 2.7264 & 3.0547 & 3.0029 & 3.1553 & 3.0506 & 2.9987 & 2.9980 & 3.7496 & 3.0920
  
                           \\ \hline
\multirow{2}{*}{BSDM \cite{BSDM}}    & AUC(PD, PF) & 0.9871 & 0.9545 & 0.9865 & 0.9370 & 0.9881 & 0.9859 & 0.9991 & 0.9863 & 0.9781
\\ %
                            & Time & 11.7277 & 11.0461 & 8.8651 & 9.2723 & 11.5979 & 11.7398 & 11.1049 & 11.7144 & 10.8835
  
                           \\ \hline
\multirow{2}{*}{OT-AD \cite{OT-AD}}    & AUC(PD, PF)  & 0.9478 & 0.9339 & 0.9859 & 0.9487 & 0.9886 & 0.9959 & 0.9909 & 0.9959 & 0.9735 
\\ %
                            & Time & 6.2187 & 8.0436 & 7.3422 & 7.4400 & 7.0764 & 7.4323 & 7.1578 & 6.1332 & 7.1055 
  
                           \\ \hline
\multirow{2}{*}{HyFuHAD}    & AUC(PD, PF)  & {\bf \red 0.9958} & {\bf \red 0.9860} & {\bf \red 0.9962} & {\bf \red 0.9735} & {\bf \red 0.9983} & {\bf \red 0.9968} & {\bf \red 0.9999} & {\bf \blue 0.9973} & {\bf \red 0.9930}
\\
                            & Time  & {1.8633} & {2.6189} & { 2.1010} & { 2.0767} & { 2.6841} & { 2.6507} & { 2.6733} & { 2.5063} & {2.3968} 
  
                           \\ \hline \hline
              
\end{tabular}
\end{table*}
%
\subsection{Ablation Study and Discussion}\label{subsec: ablation}
In Section \ref{subsubsec: ablation_MFs}, we first provide a detailed discussion of the complementarity and contributions of different fuzzy MFs in achieving effective detection performance. 
Subsequently, ablation studies presented in Section \ref{subsubsec: ablation_MCDM} demonstrate the effectiveness of incorporating Einstein Fuzzy Computing, and that the quantum MCDM can significantly improve the performance of classical MCDM.
\subsubsection{Contributions of Various MFs}\label{subsubsec: ablation_MFs}
\begin{figure}[t]
    \centering
    \includegraphics[width=1\linewidth]{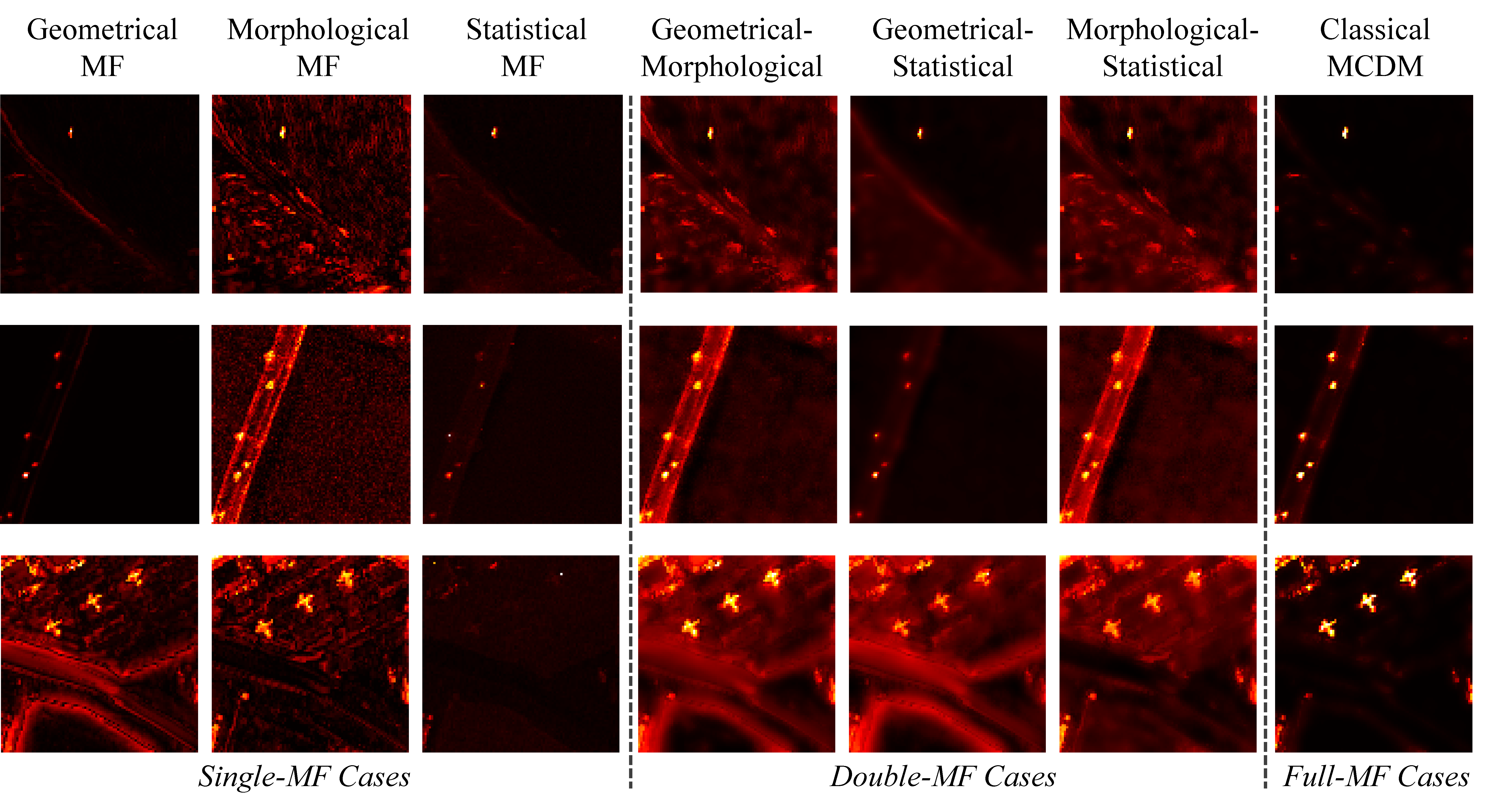}
    \caption{Qualitative comparisons between the single-MF, double-MF, and full-MF (i.e., classical MCDM) cases over the real-world Airport, Beach, and Bridge scenarios.
    As demonstrated, the full-MF case can provide superior detection performance compared with using simplified MFs.
    The corresponding quantitative evaluations and discussion are provided in Table \ref{tab: ablation study MF} and Section \ref{subsubsec: ablation_MFs}, respectively.
    }\label{fig:MF_ablation}
    \vspace{-0.3cm}
\end{figure}

\begin{table}[t]
    \caption{Ablation study of various combinations of MFs, where ``Geo.'' and ``Morph.'', and ``Stat.'' denote the abbreviations of morphological, geometrical, and statistical, respectively.   
    The marker ``\CheckmarkBold" indicates that the corresponding MF is incorporated to obtain the detection results.
    The boldfaced and underlined numbers represent the best AUC score and fastest inference time, respectively, averaged over the eight real-world datasets (see Figure \ref{fig:data}).
    The AUC scores for each combination are explicitly reported in Supplementary Table II.}
    \vspace{-0.15cm}
    \begin{center}
    \setlength{\tabcolsep}{1.2mm}
    \begin{tabular}{c c  c ||c c}
    \hline
    \hline
    \makecell[c]{Geo. MF} &\makecell[c]{Morph. MF}  &\makecell[c]{Stat. MF}  &~$\text{AUC(PD, PF)}~\!(\uparrow)$ & Time (sec.) 
     \\
    \hline
   \CheckmarkBold &   &    & 0.9666 & \underline{0.0463}
    \\
    \hline
    &  \CheckmarkBold &    &  0.9852 & 0.1141
    \\
    \hline
   &  & \CheckmarkBold  &   0.9501 & 0.0761
     \\
    \hline\hline
   \CheckmarkBold &  \CheckmarkBold &  & 0.9884 & 0.2152
     \\
    \hline
   \CheckmarkBold &   & \CheckmarkBold &  0.9804 & 0.2736
     \\
    \hline
   &  \CheckmarkBold  & \CheckmarkBold & 0.9894  & 0.2396
    \\
    \hline\hline
   \CheckmarkBold &  \CheckmarkBold  & \CheckmarkBold   & {\bf 0.9914} & 0.6736
   \\
    \hline
    \hline
    \end{tabular}
    \label{tab: ablation study MF}
    \end{center}
    \end{table} 
In the proposed HyFuHAD framework, we employ three different MFs to characterize anomalies from complementary perspectives, thereby achieving the SOTA detection performance, as reported in the last section.
This section further presents an extensive discussion of how each MF contributes to the qualitative and quantitative performance of the classical MCDM process. 
First, because the geometrical, morphological, and statistical MFs convert each hyperspectral pixel into an anomaly level (i.e., fuzzy degree), we examine them individually as preliminary detectors in this discussion, namely, the single-MF cases.
Subsequently, three different combinations of two MFs, including geometrical-morphological, morphological-statistical, and morphological-geometrical, are considered to demonstrate the improvement brought by additional information, i.e., double-MF cases.
To this end, the MCDM process (see Figure \ref{fig:classical_Fuzzy_detection}) inevitably requires slight modifications. 
Specifically, the FC step originally integrates the information from \textbf{R1} to \textbf{R3} (i.e., $\overline{\bC}$), information from \textbf{R4} (i.e., $\widetilde{\bM_{\textbf{R4}}}$), and information from \textbf{R5} (i.e., $\widetilde{\bM_{\textbf{R5}}}$), thereby deriving the integrated conclusion map $\bC$, as presented in \eqref{eq: fuzzy_conclusion}.
Once $\bC$ is obtained for the double-MF cases, the EC enhancement and spatial filtering are applied to produce the final detection result (see Figure \ref{fig:classical_Fuzzy_detection}).
In other words, the modifications are merely applied to the implementations of \textbf{R1} to \textbf{R5}. 
For the \textbf{R4} and \textbf{R5} tracks, the fuzzy AND between the three fuzzy degree maps (resp. fuzzy complement) is reduced to the AND of the two selected maps (resp. fuzzy complement), while maintaining the same procedures to obtain $\widetilde{\bM_{\textbf{R4}}}$ and $\widetilde{\bM_{\textbf{R5}}}$.
For the \textbf{R1} to \textbf{R3} tracks, the original three combinations are simplified to a single combination in the double-MF cases; however, this combination is repetitive to \textbf{R4} and provides no additional information.
Accordingly, the two selected fuzzy degree maps are treated as $\bC_1$ and $\bC_2$, and then averaged to obtain $\overline{\bC}$.
Therefore, we are able to obtain $\bC$ using $\overline{\bC}$, $\widetilde{\bM_{\textbf{R4}}}$, and $\widetilde{\bM_{\textbf{R5}}}$ by following \eqref{eq: fuzzy_conclusion}. 
Thus far, the modifications for double-MF cases have been completed. 
Finally, the full-MF case corresponds to the proposed classical MCDM framework and can be implemented as described in Section \ref{subsec: CFD}.

First, Figure \ref{fig:MF_ablation} presents several qualitative comparisons among the single-MF, double-MF, and full-MF cases.
As observed, the three MFs exhibit distinct characteristics.
For example, the morphological MF effectively highlights anomalies, whereas its background suppression capability is relatively limited.
In contrast, the statistical MF demonstrates stronger background suppression capability but insufficient anomaly detectability, thereby revealing complementary attributes to the morphological MFs.  
Regarding the geometrical MF, it exhibits scenario-specific strengths, as illustrated in Figure \ref{fig:MF_ablation}.
From these observations, the three different MFs provide additional information and complement each other.
Next, we examine several representative doblue-MF cases (see the middle columns of Figure \ref{fig:MF_ablation}) for additional validations.
For the geometrical-morphological combination on the Beach dataset, the result preserves the anomaly detectability of morphological MF, while enhancing background suppression with the guidance of geometrical MF.
On the Bridge dataset, the geometrical-statistical combination successfully integrates the detectability of geometrical MF and the background suppression capability of the statistical MF.
Furthermore, to quantitatively evaluate the advantages of incorporating additional information, the averaged AUC scores of different MF combinations over the right real-world datasets are reported in Table \ref{tab: ablation study MF}.
Due to space limitations, the dataset-wise AUC scores are provided in Supplemental Table II.
As shown in Table \ref{tab: ablation study MF}, the double-MF cases consistently outperform their corresponding single-MF counterparts in quantitative performance.
For instance, the geometrical and statistical MF achieve averaged AUC scores of 0.9666 and 0.9501, respectively.
Whereas the geometrical-statistical combination substantially improves the ACU scores to 0.9804. 
These results indicate that the incorporation of additional information for the MCDM framework can enhance detection performance.
Regarding the full-MF case (see the rightmost column of Figure \ref{fig:MF_ablation}), the qualitative performance is further enhanced by integrating the three complementary MFs.
Moreover, the full-MF case consistently achieves the best (or the second-best) ACU score across nearly all datasets (see Supplementary Table II).
Consequently, these results yield the strongest averaged performance, as summarized in Table \ref{tab: ablation study MF}.
With consistent quantitative and qualitative improvements, the necessity of three MFs has been validated.

\subsubsection{Analysis on Classical and Quantum MCDM}\label{subsubsec: ablation_MCDM}
\begin{figure}[t]
    \centering
    \includegraphics[width=1\linewidth]{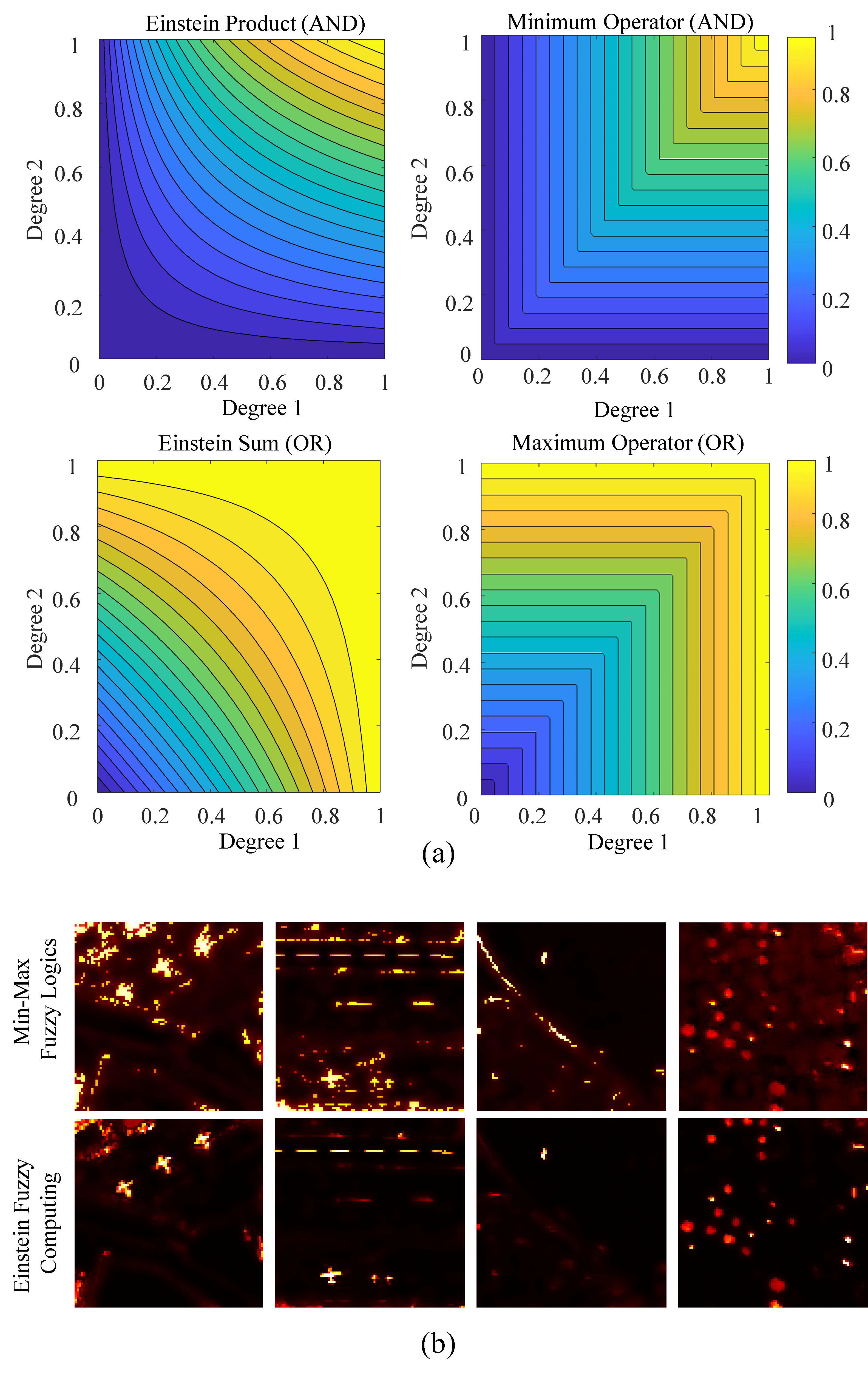}
    \caption{(a) Contour plots of fuzzy ``AND'' (i.e., Einstein product and minimum operator) and fuzzy ``OR'' (i.e., Einstein sum and maximum operator).
    (b) Qualitative comparisons between the classical MCDM implemented with Einstein fuzzy computing (proposed) and that implemented with conventional min-max fuzzy logics. 
    As demonstrated, conventional min-max operators fail to provide smooth transitions among fuzzy degrees.
    By contrast, Einstein fuzzy computing enables flexible matching of fuzzy degrees, thereby consistently yielding effective detection performance.
    }\label{fig:fuzzy_operator_ablation}
    \vspace{-0.3cm}
\end{figure}

\begin{figure}[t]
    \centering
    \includegraphics[width=1\linewidth]{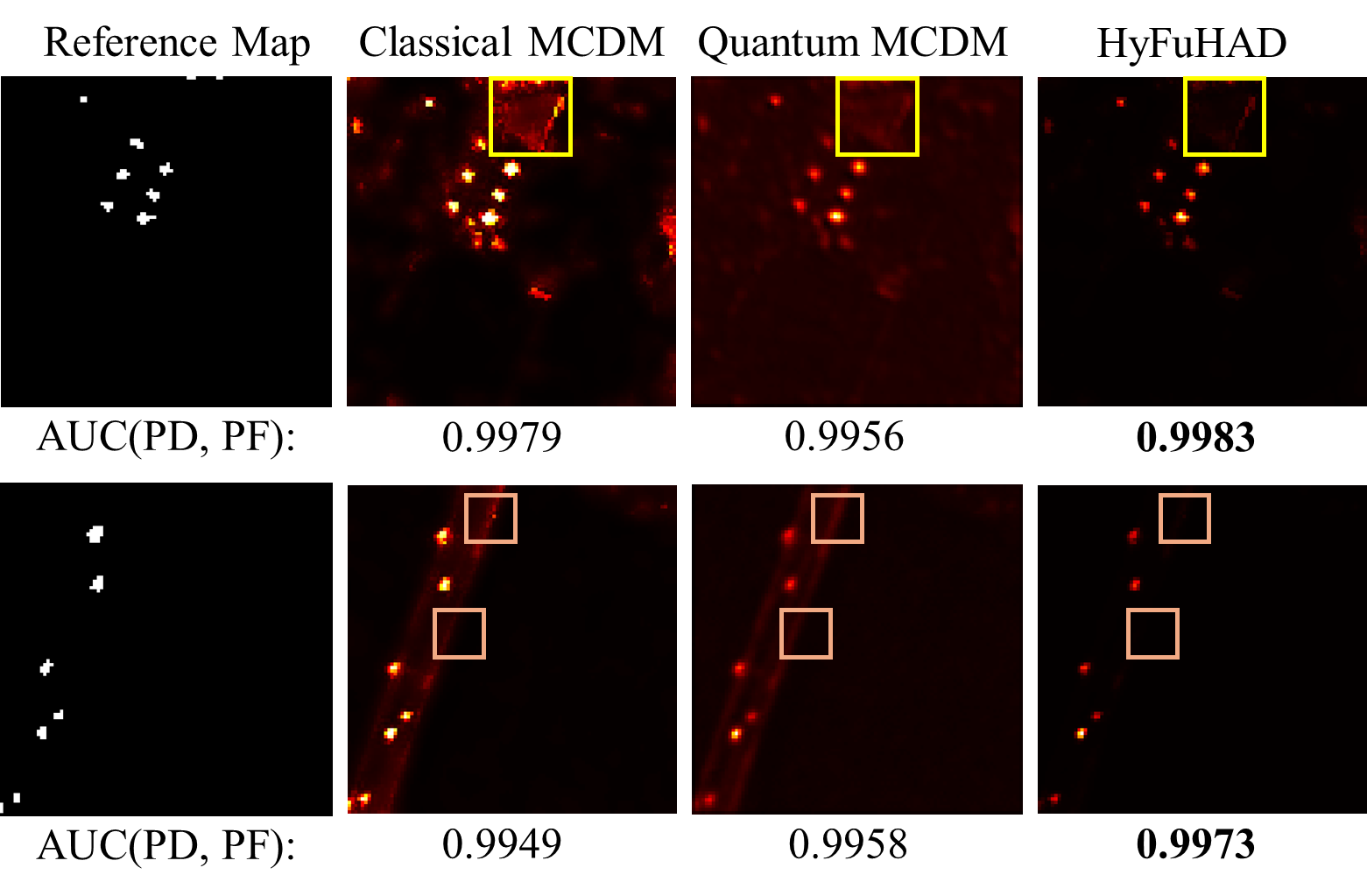}
    \caption{Qualitative comparisons between the detection results obtained by using only the classical MCDM, quantum MCDM, or the overall HyFuHAD framework over the representative real-world Urban and Bridge scenarios.}\label{fig:ablation_example}
    \vspace{-0.3cm}
\end{figure}

\begin{table}[t]
    \caption{Ablation study of the proposed HyFuHAD, where ``Classical (Min-Max)'' and ``Classical (Einstein)'' denote the classical MCDM implemented using conventional min-max fuzzy operators and Einstein fuzzy computing, respectively.   
    The marker ``\CheckmarkBold" indicates that the corresponding MCDM is incorporated to obtain the detection results.
    The boldfaced and underlined numbers represent the best AUC score and fastest inference time, respectively, averaged over the eight real-world datasets (see Figure \ref{fig:data}).
    The AUC scores for each scenario are explicitly reported in Supplementary Table III.}
    \vspace{-0.15cm}
    \begin{center}
    \setlength{\tabcolsep}{1.2mm}
    \begin{tabular}{c c  c ||c c}
    \hline
    \hline
    \makecell[c]{Classical \\ (Min-Max)} &\makecell[c]{Classical \\ (Einstein)}  &\makecell[c]{Quantum-\\ Driven}  &~$\text{AUC(PD, PF)}~\!(\uparrow)$ & Time (sec.) 
     \\
    \hline
   \CheckmarkBold &   &    & 0.9837 & \underline{0.3862}
    \\
    \hline
    &  \CheckmarkBold &    & 0.9914 & 0.6736
    \\
    \hline
   &  & \CheckmarkBold  &  0.9857 & 1.7232
     \\
    \hline
   &  \CheckmarkBold & \CheckmarkBold  &  \textbf{0.9930} & 2.3968
    \\
    \hline
    \hline
    \end{tabular}
    \label{tab: ablation study}
    \end{center}
    \end{table} 

In this section, we first present detailed comparisons between conventional min-max fuzzy operators and the Einstein fuzzy computing.
The transition behaviors of the min-max and Einstein fuzzy operators are shown in Figure \ref{fig:fuzzy_operator_ablation}(a).
In this figure, the horizontal and vertical axes correspond to two input fuzzy degrees, and the output degrees derived by a given fuzzy operator (as written in the titles) are visualized in the corresponding colors. 
Moreover, the black lines/curves represent equal degree levels, thereby forming contour plots. 
As demonstrated, the minimum and maximum operators exhibit relatively ``crisp'' behaviors, rendering the derived fuzzy degrees with limited flexibility.
To be more specific, we consider the fuzzy ``AND'' (i.e., minimum operator and Einstein product) as an example. 
Given a fixed input fuzzy degree 1 (e.g., 0.1), the contour pattern of the minimum operator indicates that the derived degrees (i.e., degree 1 ``AND'' degree 2) remain nearly unchanged, regardless of how degree 2 varies.  
In other words, when the transitional behavior exhibits right-angle contour lines, the derived degrees tend to be identical to one of the input degrees, hindering the MCDM framework from exploiting additional information.
In contrast, the Einstein product provides a much smoother degree transition, as we discussed in Section \ref{sec: intro} and Section \ref{subsec: CFD}.
Even when one input degree (e.g., degree 1) is fixed within $(0,1)$, the resulting degree can smoothly vary with another input degree when adopting the Einstein product as fuzzy AND, since its transition behavior does not exhibit perpendicular contour lines.
These observations preliminarily suggest that fuzzy operators providing smoother transition behaviors enable more informative MCDM results.

To further substantiate these observations, we replace the Einstein product (resp. Einstein sum) with a typical minimum operator (resp. maximum operator), while retaining the same settings to perform the proposed classical MCDM.
Representative results are provided as experimental evidence, as demonstrated in Figure \ref{fig:ablation_example}(b).
When the conventional min-max operators are adopted to implement fuzzy AND and OR, the detection results exhibit relatively ``crisp'' patterns [see the upper row of Figure \ref{fig:ablation_example}(b)].
In such cases, the true-detections and false-detections reveal very similar anomaly scores, which degrades the qualitative identifiability.
In contrast, the classical MCDM with Einstein fuzzy computing achieves more effective results.
As the Einstein operators can facilitate smoother degree transitions, false detections are assigned to obtain noticeably lower anomaly scores [see the bottom row of Figure \ref{fig:ablation_example}(b)].
Furthermore, the AUC scores (averaged over the eight real-world datasets) of classical MCDM with min-max operators and with Einstein fuzzy computing are reported in Table \ref{tab: ablation study}.  
We refer readers to Supplementary Table III for the explicit AUC scores of each dataset.
Overall, the ablation study substantiates the effectiveness of incorporating the Einstein fuzzy computing, which consistently yields improved AUC scores (see Supplementary Table III).
In summary, the above experiments suggest that fuzzy operators with smoother transition behaviors lead to improved detection performance for the training-free classical MCDM framework.

Subsequently, we would like to provide insights into how the quantum defuzzifier makes decisions based on the learned fuzzy features to enable better understanding.
Due to space limitations, the detailed analyses, along with feature visualizations, are presented collectively in Supplementary Figure 6.
In these analyses, the results suggest that the fuzzy feature aggregation network tends to enhance the degree of anomaly (resp. background) in distinct feature representations.
Building upon these fuzzy features, the quantum defuzzifier suppresses uncertainty within the input features while strongly enhancing the anomaly degree to achieve
effective defuzzification.
With the above insight in mind, the ablation study for classical MCDM (with Einstein operators) and quantum MCDM is presented as follows.
First, we adopt the representative examples (i.e., Urban  \uppercase\expandafter{\romannumeral 1} and Bridge), as presented in Figure \ref{fig:ablation_example}, for the following discussion.
Based on the figure, we can observe that although both classical and quantum MCDM achieve superior detections, the classical MCDM potentially has less effectiveness in suppressing certain background regions (see the box regions in Figure \ref{fig:ablation_example}).
In contrast, the quantum MCDM lacks sufficient anomaly detectability compared to the classical MCDM.
Therefore, when we simultaneously adopt the quantum and classical MCDM for the HAD task (i.e., HyFuHAD), they form a complementary detection and enable to help each other to achieve the SOTA performance (see Figure \ref{fig:ablation_example}).
To further substantiate our observation, we perform an ablation study based on the AUC score averaged over the eight real-world datasets used in our experiments.
Moreover, the AUC scores of classical MCDM, quantum MCDM, and the proposed HyFuHAD across each dataset are provided in Supplementary Table III due to space limitations.
As presented in Table \ref{tab: ablation study}, both quantum and classical MCDM are efficient and effective individually for the HAD task. 
However, the proposed HyFuHAD, which integrates both types of MCDM, consistently achieves notable quantitative improvements over using them separately across real-world datasets (see Supplementary Table III).
Accordingly, both quantum and classical MCDM are indispensable for achieving an effective HAD task.

On the other hand, a detailed discussion regarding the efficient designs of HyFuHAD is provided for further insights.
First, the trainable convolutional layers in the quantum MCDM contain only around 0.8E+3 to 1.3E+3 parameters, depending on the number of input spectral bands.
This lightweight property stems from the use of depthwise and $1\times1$ convolutions, which are significantly more efficient than typical full convolutions. 
Subsequently, the trainable hesitancy tokens are treated as part of the network parameters, with a scale that is linearly proportional to the number of pixels.
As for the QUEEN-based defuzzifier, it involves only 12 quantum neurons.
Since the classical MCDM is a non-deep learning framework, the proposed HyFuHAD solely comprises on the order of 1E+4 parameters.
To demonstrate the advantage of such a lightweight design, the computational time of the proposed framework across varying input sizes is reported, thereby analyzing its computational complexity.
The corresponding results are presented in Supplementary Figure 7 owing to space limitations.
In this analysis, the input hyperspectral structures are randomly generated with $H\times H$ pixels, as indicated along the horizontal axis.
The number of spectral bands is fixed at 224, corresponding to the AVIRIS sensor.
For reference, the linear-complexity curve (i.e., the purple curve) is additionally provided in Supplementary Figure 7.
As illustrated in the figure, the overall HyFuHAD framework, integrating both classical and quantum MCDM, demonstrates a computational growth lower than that of the reference linear complexity, implying a favorable computational efficiency. 
\subsection{Case Study}\label{subsec: Case}
\subsubsection{Large-Scale Hyperspectral Scenes}\label{subsubsec: Large_Size_Dataset}
\begin{figure}[t]
    \centering
    \includegraphics[width=1\linewidth]{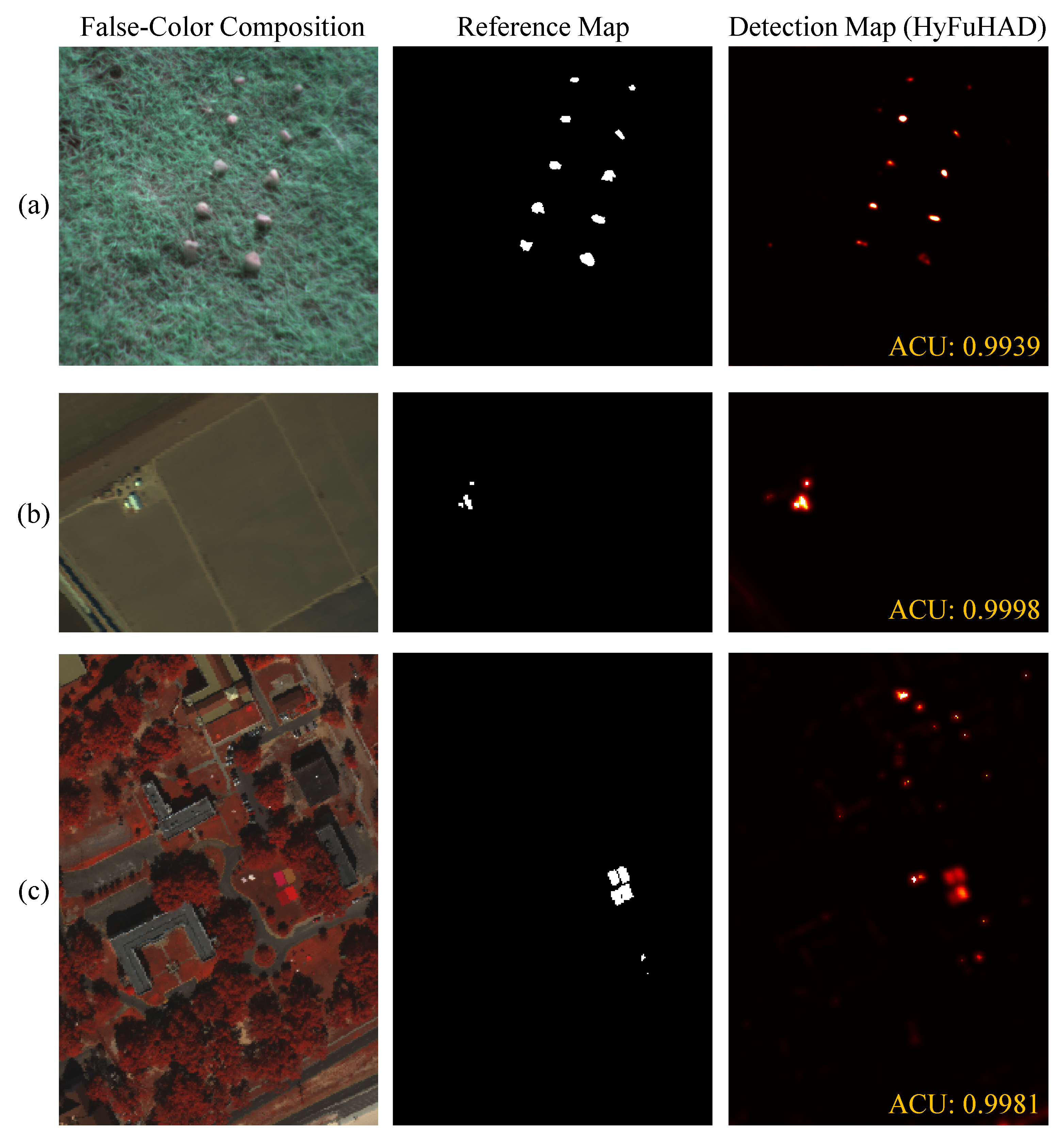}
    \caption{Case study on large-scale HAD datasets, including (a) Cri datasets with $400\times 400$ pixels, (b) Salinas datasets with $150\times 200$ pixels, and (c) MUUFL datasets with $325\times 220$ pixels. 
    As demonstrated, the proposed HyFuHAD yields effective detection performance across diverse large-scale datasets.
    More details are presented in Section \ref{subsubsec: Large_Size_Dataset}.
    }\label{fig:large_size_data}
    \vspace{-0.3cm}
\end{figure}
In this section, we further employ the publicly available HAD datasets with large spatial dimensions to validate the practical applicability of our proposed HyFuHAD.
Brief introductions of these large-scale HAD datasets are provided as follows.
The Cri dataset \cite{9246710}, as shown in Figure \ref{fig:large_size_data}(a), was captured by the Nuance Cri hyperspectral sensor with a 10 nm spectral resolution over an artificial scene at Wuhan University, China.
The HSI contains a total of $400\times 400$ pixels with $C=46$ spectral bands, where $1254$ hyperspectral pixels are labeled as anomalies.
Subsequently, the Salinas dataset [see Figure \ref{fig:large_size_data}(b)] was acquired by AVIRIS sensors over Salinas Valley, CA, USA \cite{SuperRPCA}.
This dataset consists of $150\times 200$ pixels with $204$ spectral bands, where a total of $46$ pixels are identified as anomalies (i.e., farmhouse).   
The MUUFL dataset \cite{MUUFL} illustrated in Figure \ref{fig:large_size_data}(c) was captured over the University of Southern Mississippi, USA, by the CASI-1500 hyperspectral sensor.
The spatial and spectral dimensions are $325\times 220$ and $64$, respectively.
Among these pixels, $269$ pixels identified as cloth targets on the campus are labeled as anomalies.
For the large and complex MUUFL dataset, we set the number of training iterations to 45 and retain training epochs to 30 for the Cri and Salinas datasets. 

The qualitative and quantitative performance are collectively presented in Figure \ref{fig:large_size_data}.
In the detection result of the Cri dataset [see Figure \ref{fig:large_size_data}(a)], the proposed HyFuHAD effectively identifies the artificial anomaly targets in the upper-right region.
The qualitative evaluation demonstrates the strong background suppression capability and anomaly detectability of our HyFuHAD.
Even after substantially enhancing the intensity of the resulting maps, the observed patterns remain only anomalies, rather than exhibiting both the background and the targets.  
Moreover, the superior AUC score of 0.9939 quantitatively validates the strong detection performance. 
On the other hand, as illustrated in Figure \ref{fig:large_size_data}(b) and Figure \ref{fig:large_size_data}(c), detection results on the Salinas and MUUFL datasets consistently support the effectiveness of HyFuHAD.
In these amplified representations, the proposed HyFuHAD identifies the anomalies with clearly distinguishable intensities for the Salinas dataset, thereby yielding a nearly perfect AUC score (i.e., 0.9998).
Although a small number of background components are suppressed less effectively in the upper-right region of the MUUFL datasets, HyFuHAD still achieves an overall AUC score of 0.9981.
This observation suggests that our HyFuHAD is able to identify targets with distinguishable anomaly levels.
In summary, the above experimental results show that, even when the spatial dimension of input HSIs increases significantly, the proposed HyFuHAD remains effective in detecting targets with abnormal spectral characteristics.

\subsubsection{Oil Spill Detection}\label{subsubsec: Oil}
\begin{figure}[t]
    \centering
    \includegraphics[width=1\linewidth]{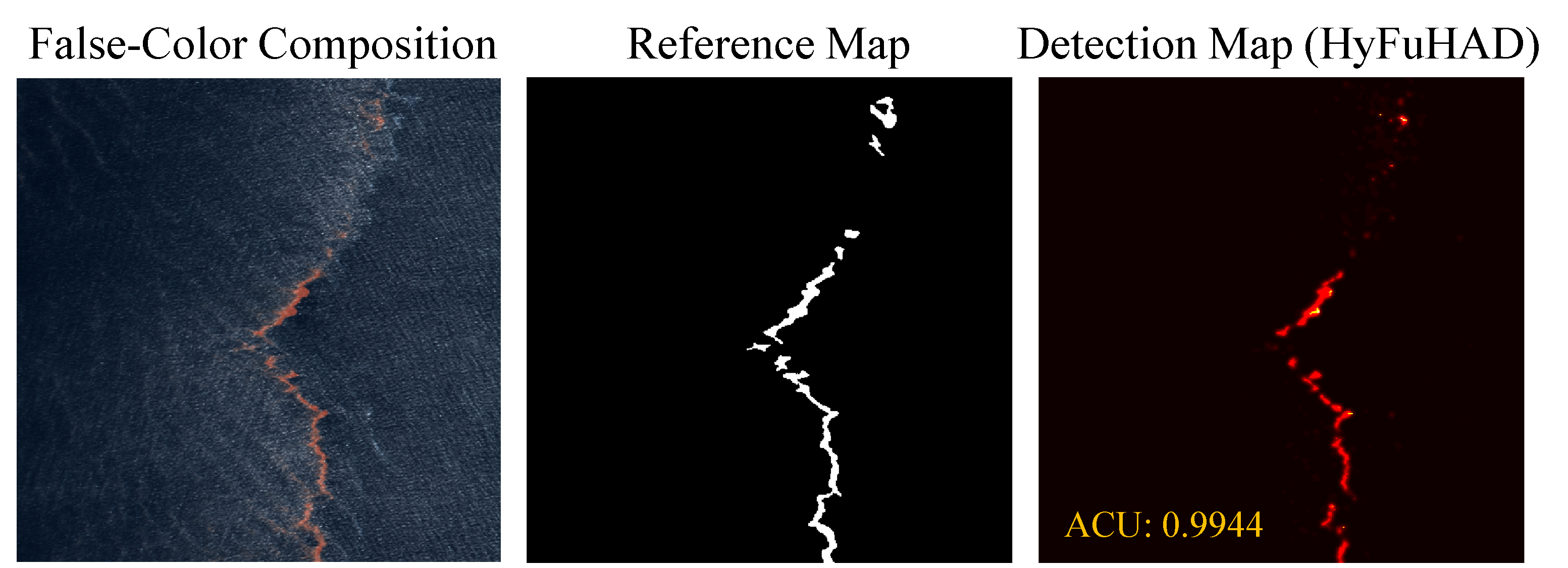}
    \caption{Hyperspectral oil spill detection (HOSD) using the proposed HyFuHAD framework.
    The investigated region has a spatial size of $512\times 512$ pixels and was acquired by the AVIRIS sensor over the Gulf of Mexico on May 17, 2010.  
    As oil spills exhibit spectral characteristics significantly deviated from the background ocean  (i.e.,  anomalies), we leverage the proposed HyFuHAD as a fully unsupervised HOSD alternative and successfully identify oil leakages.
    More details are presented in Section \ref{subsubsec: Oil}.
    }\label{fig:oill_spill_detection}
    \vspace{-0.3cm}
\end{figure}
In this section, we apply the proposed HyFuHAD framework to a critical real-world application, namely oil spill detection (OSD), to demonstrate its practical applicability.
In recent years, oil-spill accidents have been frequently reported during the marine oil exploitation and transportation.
Marine oil spills that are not adequately and timely remediated pose a severe threat to natural environments and pollute marine ecosystems.  
With the modern RS technique, an advanced hyperspectral imaging system (e.g., the AVIRIS sensor) can provide rich spatial and spectral information for marine oil spill monitoring, i.e., hyperspectral OSD (HOSD) \cite{haut2023cloud}.
Nevertheless, developing supervised machine learning frameworks or deep networks for HOSD may face several challenges, including limited availability of large-scale training data and manual annotations.
We observe that,  in a large-scale marine oil spill accident, the clean ocean (considered as background) and the oil-polluted regions exhibit significantly different spectral characteristics, which closely adhere to the properties of a hyperspectral anomaly.
As a result, employing a fully unsupervised HAD framework as an alternative exhibits great potential for addressing the challenging HOSD task.  
In the following experiments, we demonstrate the feasibility of using the proposed HyFuHAD framework as an unsupervised alternative for HOSD. 

To this end, we employ the publicly available HOSD database \cite{HOSD_Data} acquired by the AVIRIS sensor on May 17, 2010, over the Gulf of Mexico (GM).
To ensure spectral fidelity, the water-vapor absorption bands \cite{HyperQUEEN,PRIME}, including bands 1–10, 104–116, 152–170, and 215–224, are excluded before experimental analysis.
The investigated area, shown in Figure \ref{fig:oill_spill_detection}, comprises $512\times 512$ pixels with a spatial resolution of 7.6 m and corresponds to the most severe marine oil pollution events in US history.
The oil spill accident was caused by the explosion of the British Petroleum (BP) Deepwater Horizon Macondo oil well drilling platform \cite{Oil_Spill}.
As demonstrated in Figure \ref{fig:oill_spill_detection}, the oil-leakage area exhibits significantly deviated spectral characteristics compared with the clean ocean background. 
Accordingly, when we apply the proposed HyFuHAD framework to the HOSD task, the resulting intensity-amplified detection map (see Figure \ref{fig:oill_spill_detection}) reveals a consistent oil-leakage pattern that closely matches the corresponding reference map.
With accurate identification of the oil-leakage area, the detection map achieves an AUC score of 0.9944, despite HyFuHAD being originally developed for the HAD task.  
The experiment not only substantiates the feasibility of performing HOSD using an unsupervised HAD framework but also demonstrates the effectiveness of the proposed HyFuHAD.

\section{Conclusion}\label{Conclusion}
To address the challenging RS task of HAD, this study develops the HyFuHAD framework by integrating the classical and quantum fuzzy MCDMs (see Figure \ref{fig: overall_famework}).
In the classical MCDM, we employ three key HAD properties, including morphological, geometrical, and statistical, to customize the fuzzification step for the HAD task.  
These properties are not only informative but also mutually complementary (see Section \ref{subsubsec: ablation_MFs}), thereby leading to effective classical detection results.
During the FM and FI steps (see Figure \ref{fig:classical_Fuzzy_detection}), Einstein fuzzy operators serve as the fuzzy ``AND'' and ``OR'' to implement the five HAD-based fuzzy rules.
Einstein fuzzy computing facilitates smoother degree transition behaviors compared to widely adopted fuzzy operators (e.g., minimum and maximum operators).
As experimentally validated in Section \ref{subsubsec: ablation_MCDM}, the proposed classical MCDM implemented using the Einstein fuzzy computing consistently outperforms its counterparts using conventional fuzzy operators.

To further model complex real-world scenarios, we unfold the five HAD-based fuzzy rules into a deep fuzzy feature aggregation network to learn a discriminative deep fuzzy representation.
Furthermore, inspired by the extended fuzzy set, we introduce trainable hesitancy tokens into the deep fuzz space to characterize the ambiguity between anomaly and background, thereby providing enhanced flexibility and robustness against noise corruptions (see Supplementary Figure 5).
Motivated by the widespread  QUEEN in the RS field, we incorporate the advanced lightweight QUEEN, which comprises solely 12 quantum neurons, as a complex-valued defuzzifier to obtain the quantum-based detection result.
The resulting quantum-empowered decision framework is known as the quantum MCDM (see Figure \ref{fig:Fuzzy_detection}).
Experimental results demonstrate that the proposed HyFuHAD achieves the SOTA performance by fusing the quantum and classical information.
Extensive ablation studies further verify the necessity of integrating both quantum and classical MCDM.

Beyond the widely used HAD datasets, which have relatively small spatial dimensions, several publicly available large-scale HAD datasets (i.e., Cri, Salinas, and MUUFL datasets) are additionally employed to evaluate the practical applicability of the proposed HyFuHAD framework.
In these case studies (see Section \ref{subsec: Case}), the proposed HyFuHAD framework consistently achieves effective qualitative and quantitative detection performance across all scenarios.
Moreover, attributed to the efficient design of both classical and quantum MCDM, the overall HyFuHAD framework exhibits a computational growth lower than that of linear complexity.
This favorable computational efficiency greatly enhances the practical utility of HyFuHAD for large-scale remotely sensed HSIs.
In addition, we evaluate the proposed HyFuHAD as an unsupervised HOSD alternative.
As shown in Figure \ref{fig:oill_spill_detection}, our HyFuHAD again achieves an effective AUC score of 0.9944.
These experimental analyses collectively substantiate the practical effectiveness and applicability of the proposed HyFuHAD framework.
%
\renewcommand{\thesubsection}{\Alph{subsection}}
\bibliographystyle{IEEEtran}
\bibliography{ref}

\begin{IEEEbiography}[{\resizebox{0.9in}{!}{\includegraphics[width=1in,height=1.25in,clip,keepaspectratio]{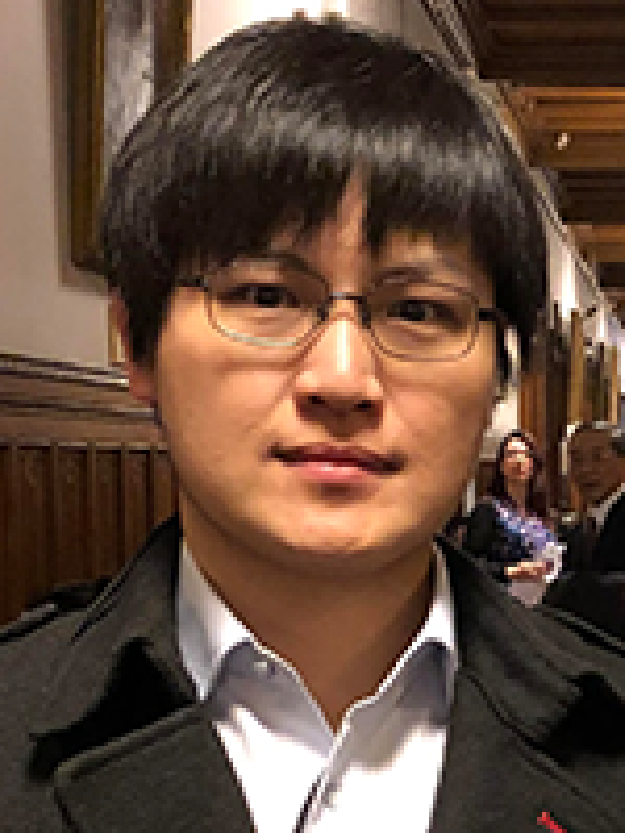}}}]
	{\bf Chia-Hsiang Lin}
(S'10-M'18-SM'24)
received the B.S. degree in electrical engineering and the Ph.D. degree in communications engineering from National Tsing Hua University (NTHU), Taiwan, in 2010 and 2016, respectively.
From 2015 to 2016, he was a Visiting Student of Virginia Tech,
Arlington, VA, USA.

He is currently a Professor with the Department of Electrical Engineering,
National Cheng Kung University (NCKU), Taiwan, and also serves as a Technical Director of Smart Sensing \& Systems Technology Center, Industrial Technology Research Institute (ITRI).
Before joining NCKU, he held research positions with The Chinese University of Hong Kong, HK (2014 and 2017),
NTHU (2016-2017),
and the University of Lisbon (ULisboa), Lisbon, Portugal (2017-2018).
He was an Assistant Professor with the Center for Space and Remote Sensing Research, National Central University, Taiwan, in 2018, a Visiting Professor with ULisboa, in 2019, and a Visiting Professor with Texas A\&M University, USA, in 2025.
His research interests include network science,
quantum computing,
convex geometry and optimization, blind signal processing, and imaging science.

Dr. Lin received the Emerging Young Scholar Award (The 2030 Cross-Generation Program) from National Science and Technology Council (NSTC), from 2023 to 2027,
the Future Technology Award from NSTC, in 2022,
the Outstanding Youth Electrical Engineer Award from The Chinese Institute of Electrical Engineering (CIEE), in 2022,
the Best Young Professional Member Award from IEEE Tainan Section, in 2021,
and the Prize Paper Award from IEEE Geoscience and Remote Sensing Society (GRS-S), in 2020.
He received the Ministry of Science and Technology (MOST) Young Scholar Fellowship, together with the EINSTEIN Grant Award, from 2018 to 2023.
In 2016, he was a recipient of the Outstanding Doctoral Dissertation Award from the Chinese Image Processing and Pattern Recognition Society and the Best Doctoral Dissertation Award from the IEEE GRS-S.
\end{IEEEbiography}

\begin{IEEEbiography}[{\resizebox{0.9in}{!}{\includegraphics[width=1in,height=1.25in,clip,keepaspectratio]{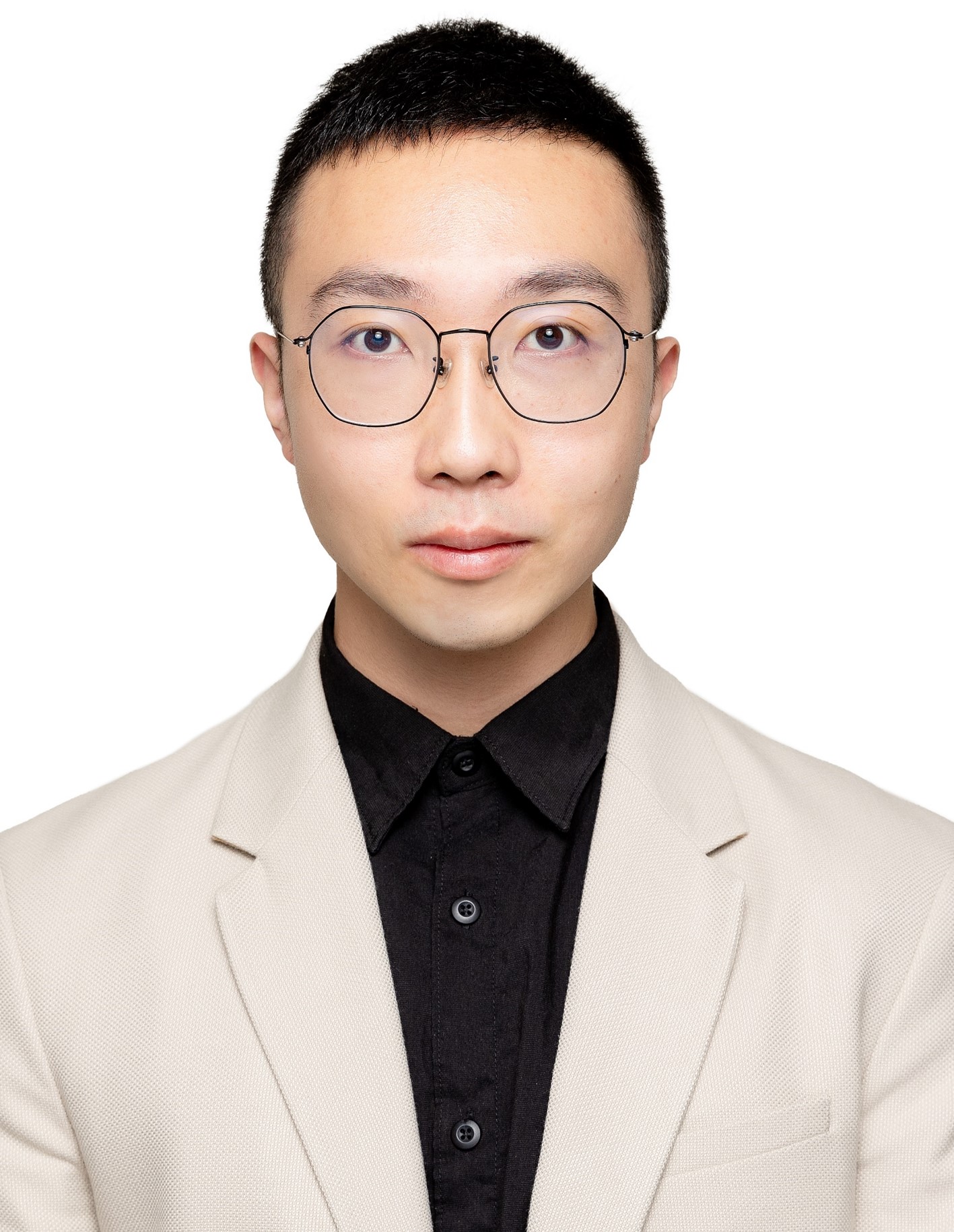}}}]
	{\bf Si-Sheng Young}
(S'23) is currently a Ph.D. student with the Intelligent
Hyperspectral Computing Laboratory (IHCL), Department of Electrical Engineering, National Cheng Kung University (NCKU), Tainan, Taiwan. 

In 2023, he received the Merit Award from The Grand Challenge ``Computing for the Future", Miin Wu School of Computing, NCKU, as well as the highly competitive ``Pan Wen Yuan Scholarship" from the Industrial Technology Research Institute (ITRI), Hsinchu, Taiwan.
In 2024, he received a highly competitive ``Scholarship Pilot Program to Cultivate Outstanding Doctoral Students" from the National Science and Technology Council (NSTC), Taiwan.
His research interests include convex optimization, deep learning, anomaly detection, data fusion, and imaging inverse problems.
\end{IEEEbiography}

\begin{IEEEbiography}[{\resizebox{0.9in}{!}{\includegraphics[width=1in,height=1.25in,clip,keepaspectratio]{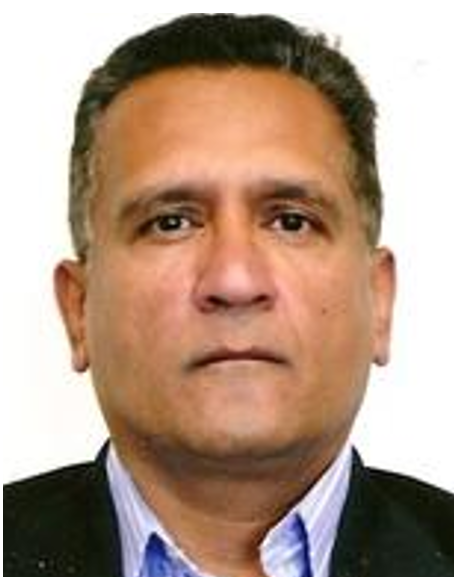}}}] {\bf Reza Langari} (Senior Member, IEEE)  received the B.Sc., M.Sc., and Ph.D. degrees from the University of California at Berkeley, CA, USA, in 1981, 1983, and 1991, respectively.

He was with Measurex Corporation from 1984 to 1995, Integrated Systems Inc., from 1985 to 1986, and Insight Development Corporation from 1987 to 1989, prior to starting his academic career with Texas A\&M University, College Station, TX, USA, in September 1991. 
He has since held research positions at the National Aeronautics and Space Administration Ames Research Center, the Rockwell International Science Center, the United Technologies Research Center, and the U.S. Air Force Research Laboratory.
He has coauthored \textit{Fuzzy Logic: Intelligence, Control and Information} (Prentice Hall, 1999) and \textit{Measurement and Instrumentation} (Elsevier, Second Edition, 2011, 2015). 

Dr. Langari was elected a Fellow of the American Society of Mechanical Engineers (ASME) in 2009.
He also received the Lifetime Achievement Award from the World Automation Congress in 2022.
His current research interests include computational intelligence, with application to mechatronic systems, industrial automation, and automated vehicles. 
He has served as an Associate Editor for IEEE TRANSACTIONS ON FUZZY SYSTEMS, the IEEE TRANSACTIONS ON VEHICULAR TECHNOLOGIES, and \textit{ASME Journal of Dynamic Systems, Measurement, and Control}.
He currently serves as the Editor-in-Chief for the \textit{Journal of Intelligent and Fuzzy Systems}.
\end{IEEEbiography}

\end{document}